\documentclass[12pt]{amsart}

\pdfoutput=1

\usepackage{amsmath} 
\usepackage{amssymb} 
\usepackage{bbm}
\usepackage{graphicx} 
\usepackage{caption}
\usepackage{subcaption}
\usepackage{color} 
\usepackage[open=true]{bookmark}
\usepackage{url}
\hypersetup{ 
  colorlinks=true, 
  citecolor=red, 
  filecolor=black, 
  linkcolor=cyan, 
  urlcolor=magenta 
}

\usepackage[english]{babel} 
\theoremstyle{plain} 
\newtheorem{theorem}{Theorem}
 
\newtheorem*{definition}{Definition}

\newtheorem{conjecture}{Conjecture} 

\newtheorem*{remark}{Remark} 
\newtheorem*{statement}{Statement} 

\usepackage{enumitem}

\newcommand{\bT}{{\mathbb T}}
\newcommand{\bS}{{\mathbb S}}
\newcommand{\bR}{{\mathbb R}}
\newcommand{\bZ}{{\mathbb Z}}

\newcommand{\mA}{\mathcal{A}}
\newcommand{\mG}{\mathcal{G}}
\newcommand{\mH}{\mathcal{H}}
\newcommand{\mM}{\mathcal{M}}

\newcommand{\mQ}{\mathcal{Q}}

\newcommand{\lra}[1]{\langle #1 \rangle}

\newcommand{\rmd}{\mathrm{d}}

\newcommand{\rhof}{\rho_{\mathrm{f}}}
\newcommand{\rhom}{\rho_{\mathrm{m}}}
\newcommand{\rhoc}{\rho_{\mathrm{c}}}
\newcommand{\tw}{\tau_{\mathrm{w}}}
\newcommand{\tb}{\tau_{\mathrm{b}}}

\begin{document} 

\title[Mathematical billiards \& statistical physics]
{What mathematical billiards teach us about statistical physics?}

\author[P~B\'alint]{P\'eter~B\'alint}
\address{Institute of Mathematics\\Budapest University of Technology
  and Economics and\\
  BME Stochastics Research Group of the Office for Research Groups\\
  Attached to Universities and Other Institutions (ELKH TKI)\\
  Egry J\'ozsef u. 1\\Budapest 1111\\Hungary}
\email{pet@math.bme.hu}
\author[T~Gilbert]{Thomas~Gilbert}
\address{  Centre d'\'Etude des Ph\'enom\`enes Non-Lin\'eraires et Syst\`emes
  Complexes\\Universit\'e Libre  de Bruxelles\\Campus Plaine C.~P.~231\\1050
  Bruxelles\\Belgium}
\email{thomas.gilbert@ulb.ac.be}
\author[D~Sz\'asz]{Domokos~Sz\'asz}
\address{Institute of Mathematics\\Budapest University of Technology
  and Economics\\Egry J\'ozsef u. 1\\Budapest 1111\\Hungary}
\email{domaszasz@gmail.com}
\author[I~P~T\'oth]{Imre~P\'eter~T\'oth}
\address{Institute of Mathematics\\Budapest University of Technology
  and Economics and\\
  BME Stochastics Research Group of the Office for Research Groups\\
  Attached to Universities and Other Institutions (ELKH TKI)\\
  Egry J\'ozsef u. 1\\Budapest 1111\\Hungary}
\email{mogy@math.bme.hu}
\thanks{PB, DSz, IPT are supported by Hungarian National Foundation
  for Scientific Research grants Nos. K 104745 and K 123782 and by the
  OMAA-103\"ou6 project. } 
\thanks{TG is financially supported by the (Belgian) FRS-FNRS}
\begin{abstract}
  We survey applications of the theory of hyperbolic (and to a lesser
  extent non hyperbolic) billiards to some fundamental problems of
  statistical physics and their mathematically rigorous derivations in
  the framework of classical Hamiltonian systems.
\end{abstract}

\maketitle

\smallskip
\noindent \textbf{Keywords.}
Sinai billiards, Wind--tree models, Ergodic hypothesis, Brownian
motion, Fourier's law of heat conduction.



\tableofcontents

\section*{Introduction}
\label{sec:intro}

The pursuit of mathematical rigour has arguably pervaded physical
sciences from as early as Galileo's times. Statistical mechanics,
which was founded in the second half of the 19\textsuperscript{th}
century by the likes of Rudolf Clausius (1822--1888), James Clerk
Maxwell (1831--1879), Ludwig Boltzmann (1844--1906), and Josiah
Willard Gibbs (1839--1903), has ever since remained at the forefront
of this endeavour.

However, as Landau and Lifshitz wrote back in 1937--1939 \cite[Preface
to early Russian editions]{LL},
\begin{quote}
  It is a fairly widespread delusion among physicists that statistical
  physics is the least well-founded branch of theoretical
  physics. Reference is generally made to the point that some of its
  conclusions are not subject to rigorous mathematical proof; and it is
  overlooked that every other branch of theoretical physics contains
  just as much non-rigorous proofs, although these are not regarded as
  indicating an inadequate foundation for such branches.
\end{quote}

The same authors were careful to further warn us that, in general,
``mathematical rigour is not readily attainable in theoretical
physics.'' And, indeed, statistical physics may have helped to shape
their view.  

To be sure, the foundations of statistical mechanics were, from the
onset, the subject of intense debate. Foremost among the issues that
came under scrutiny is the ergodic hypothesis \cite{EE12}, which was
understood to be a 
preliminary to showing that gas molecules behave in a stochastic
fashion.  Roughly speaking, an ergodic system\footnote{Herein,
  ergodicity is meant as a property of a probability measure, and,
  unless explicitly stated, not of an infinite one.} must be such that
almost all initial conditions yield trajectories that fill evenly the
subset of the phase space compatible with its conservation
laws\footnote{Somewhat 
  more precisely, ergodicity means that, while in equilibrium and over long
  periods of time, the time spent by a system in some region of the
  phase space of microstates with compatible energy is proportional to
  the volume of this region.}.  There was,
however, compelling evidence that \textit{Hamiltonian systems} are
generically not ergodic, which was 
ultimately encapsulated in the work of Markus and Meyer \cite{MM74}; 
see section \ref{sec:ergo} for further details.  The consensus among
physicists and mathematicians alike thus emerged that ergodic 
Hamiltonian systems are the exception rather than the rule, see
e.g.~\cite[p 137]{Smale1980} or \cite[Section 4]{Berkovitz2006}, and led 
some authors to question, paraphrasing Wigner \cite{W60}, the
``unreasonable effectiveness of statistical 
mechanics'' \cite[p 75]{Gr87}. The subject is indeed rife with
controversy \cite{Earman1996}.

The theory of \textit{hyperbolic billiards}, initiated by the seminal
1963 and 1970 works of Sinai \cite{S63,S70}, provides a singular
exception to this unpleasant state of affairs, which has been hailed
as a milestone in the development of ergodic theory; see \cite{S19}. To
this day, and apart from the gas of hard balls, only a few examples of
ergodic Hamiltonian systems have been identified \cite{K76, KM81, K87,
  DL91}. Although generic Hamiltonian systems with a fixed number of
degrees of freedom should indeed not be expected to be ergodic,
it must be said that Boltzmann's ergodic hypothesis refers to the
so-called thermodynamic limit, where the number of particles grows to  
infinity. It remains unclear whether the fulfilment of this
form of the ergodic hypothesis does require the type of hard core
interaction exhibited by billiards or is in fact applicable to Hamiltonian
systems with smooth interactions. The conjectures drawn in Subsection
\ref{ssec:infty} reflect similar expectations.

Although the singularities incurred by billiards lead to a number of
conceptual and technical difficulties, Sinai's unique achievement lied
primarily in creating the methods to handle them. Sinai's theory of
hyperbolic billiards thus opened new perspectives for
investigating the foundations of statistical physics. The main
objective of this paper is to collect a few key problems of
statistical physics to which billiard models can be successfully
applied. When it comes to the derivation of fundamental laws in a 
Hamiltonian framework, billiard models---in spite of their
singularities---are mathematically more easily tractable than other
types of interactions. 

The applications of billiard theory to
statistical physics we wish to illustrate are the following:
\begin{enumerate}
\item Boltzmann's ergodic hypothesis;
\item Brownian motion;
\item Fourier's law of heat conduction.
\end{enumerate}
They will be addressed in separate sections.

Concerning item (2) above, we should remark that Einstein's 1905 proof
\cite{E05}---however ingenious it was---was far from being  
rigorous\footnote{Its conditions do not actually hold, but this fact,
  of course, does not diminish its utmost radical novelty and scientific
  significance; see the excellent review \cite{D06}.}. The important
and interesting feature we wish to illustrate is that the derivation
of the macroscopic law of Brownian motion\footnote{%
  \emph{Brownian motion} generically refers to the diffusive
  transport process of a tagged particle, which therefore implies
  the existence of the underlying transport coefficient, whose
  expression can be inferred through the Green-Kubo formula, or
  equivalent techniques; see, e.g.,~\cite{G98}. 
},
starts from the deterministic dynamics at the so-called microscopic
level, which 
therefore entails accounting for breaking the time-reversal symmetry
of the law of motion at the micro-scale to justify the irreversible
diffusive behaviour at the macro-scale. Mathematically speaking, this
derivation relies on the appropriate control of correlation decay. 

We should also note that items (2) and (3) both deal with
diffusive processes. However, in the former, the 
focus is on mass transport of a tagged particle, whereas, in the
latter, energy is being transported. From a mathematical point of view,
these problems are quite different; while the former can be modeled with a
low-dimensional system, the latter ultimately results from the
interactions of infinitely many particles. It should further be noted that the
ideas discussed here are still under progress, even though, on a
physical level, they are already rather well understood. 

Beyond these three items, we also formulate a few open problems we
consider interesting and draw up conjectures that we  find compelling.

On the occasion of Sinai's award of the 2014 Abel prize, several
surveys were dedicated to his numerous scientific achievements; see
the volume \cite{A19}, which includes three contributions devoted to
the theory of hyperbolic billiards \cite{B19,S19,Sz19} and that cover 
a wide range of topics in detail. We will therefore refer to these
surveys whenever suitable and go into further details only where we
feel more recent developments so warrant.  Moreover, we do not treat
important problems such as quantum billiards, the Boltzmann-Grad
limit (with a couple of exceptions), or the Boltzmann equation.

We make one additional remark which pertains to the contents of our
survey. There are mathematical or physical constructions that are
isomorphic to  billiard models. Just to name some of them, consider
the Lorentz gas, the Rayleigh gas, or the wind--tree gas. If results
related to them are obtained by the methods of hyperbolic billiards,
we consider them relevant to the present survey. Otherwise our selection
will be somewhat arbitrary; we mention topics that are connected to
some interesting problem or phenomenon related to billiards.

About the structure of this paper: The models to be treated are
introduced in Section \ref{sec:models}, with
relevant notations. In Sections  
\ref{sec:ergo}, \ref{sec:diff} and \ref{sec:fourier}, the three
problems mentioned above are respectively discussed. In 
Appendix, we recall two results related to two Conjectures presented
in Subsection \ref{ssec:infty}.

\section{Models}
\label{sec:models}

We introduce below the different models that will be treated in the
sequel. We start with the general definition of billiards in
Subsection \ref{ssec:bill}, and present several useful
definitions. Lorentz processes and gases are 
subsequently defined in Subsections \ref{sec:Lproc} and \ref{sec:Lgas}
respectively. For our purpose, the former refers to a single particle
on a dispersing billiard table and the latter to a countable number of
copies of it. In Subsection \ref{ssec:wind}, examples of
non-dispersing billiards are presented. They are generally referred to
as wind--tree models. We then turn to higher-dimensional billiards,
beginning with general hard ball systems in Subsection
\ref{ssec:hardballs}. A class of such systems with spatial ordering is
presented in Subsection \ref{ssec:spatial}. We finish with the 
Rayleigh gas in Subsection \ref{ssec:rayleigh}.

\subsection{Billiards}
\label{ssec:bill}

As far as notations go, we mainly follow \cite{ChM06} for planar
billiards and \cite{BChSzT03} for multidimensional ones.

Billiards are defined in Euclidean domains bounded by a finite number
of smooth boundary pieces.  For our purpose a \textit{billiard} is a
dynamical system describing the motion of a point particle in a
connected, compact domain $\mQ \subset \bT^{d} = \bR^{d} /
\bZ^{d}$. In general, the boundary $\partial \mQ$ of the domain is
assumed to be piecewise $C^3$-smooth, i.e.~there are no corner points;
if $0<J < \infty$ is the number of such pieces, we can write $\partial
\mQ = \cup_{1\leq \alpha\leq J}\,\partial \mQ_\alpha$. Connected
components of $ \bT^{d} \setminus \mQ$ are  
called \textit{scatterers}. Motion is uniform inside $\mQ$ and specular
reflections take place at the boundary $\partial \mQ$; in other words,
a particle propagates freely until it collides with a scatterer, where
it is reflected elastically, i.e.~following the classical rule that the
angle of incidence be equal to the angle of reflection.  

Since the absolute value of the velocity is a first integral of
motion, the phase space of our billiard is defined as the product of
the set of spatial configurations by the $(d-1)$-sphere, $\mM=\mQ\times
\bS_{d-1}$, which is to say that every phase point $x\in \mM$ is of the
form $x=(q,\,v)$, with $q\in \mQ$ and $v\in \bR^{d}$ with norm
$|v|=1$. 
According to the reflection rule, $\mM$ is subject to identification
of incoming and outgoing phase points at the boundary
$\partial\mM=\partial\mQ\times \bS_{d-1}$. 
The billiard dynamics on $\mM$ is called the \textit{billiard flow}
and denoted by $S^{t}: t \in (-\infty, \infty)$, where $S^{t}: \mM \to
\mM$. The set of points defined by the trajectory going through
$x\in\mM$ is denoted $S^{\bR}x$. The smooth, invariant probability
measure of the billiard flow, $\mu$ on $\mM$, also called the
Liouville measure, is essentially the product of Lebesgue measures on
the respective spaces,
i.e.~$\rmd\mu= {\rm const.}\,  \rmd q \, \rmd v$, where the constant
is $(\textrm{vol} \,\mQ \ \textrm{vol}\, \bS_{d-1})^{-1}$. 

For later reference we recall the time-discrete \textit{collision map}
(or billiard map, or Poincar\'e section map) of the billiard flow. Let
us denote by $\partial \mM = \partial \mQ \times \bS_{d-1}^+$ the set 
of all phase points with spatial coordinates at the boundary of a
scatterer and velocities pointing outwards
(here $\bS_{d-1}^+$ refers to the corresponding hemisphere).  For a
point $x \in \mM$, let 
$\tau(x) = \min \{s > 0: S^{s}(x) \in \partial \mM\}$ denote the first
hitting time on the billiard boundary. Then for $(\xi, \,v)
\in \partial \mM$, $\tau$ is the first return time to the boundary and
the collision map  $T: \partial \mM \to \partial \mM$ is defined via
$T(\xi, \,v) = S^{\tau}(\xi, \,v)$. The natural invariant measure for 
the collision map is then $\rmd\nu(\xi, \,v) = 
\text{const.}\, \lra{n(\xi), v} \, \rmd\xi \, \rmd v$ where $n(\xi)$
is the unit normal vector of the boundary $\partial \mQ$ at $\xi
\in \partial \mQ$, directed inward $\mQ$, $\lra{\cdot, \cdot}$ is the scalar
product, and the constant is $(d-1)\,(\textrm{vol} \,\partial \mQ \
\textrm{vol}\, \bS_{d-2})^{-1}$.  

We end with the following two sets of definitions.

\begin{definition}[Dispersing and semi-dispersing Billiards]
  We say that a billiard is \emph{dispersing}
  (resp. \emph{semi-dispersing}) if its smooth 
  boundary pieces, i.e.~the scatterers, are strictly convex
  (resp. convex) when viewed from inside $\mQ$. Because of these
  \emph{convexity properties}, semi-dispersing billiards, whose
  pre-eminent examples are hard ball systems in parallelepipeds or on
  tori, exhibit different degrees of \emph{hyperbolicity}. In this
  paper, we generally refer to planar dispersing billiards as
  \emph{Sinai billiards}; see~\cite{S70}. Sinai was also responsible
  for initiating the study of higher-dimensional dispersing and
  semi-dispersing billiards; see in particular \cite{Chernov1982, SCh87}. 
\end{definition}

\begin{definition}[Infinite and finite horizons]
  Collision--free orbits are a distinctive features of some billiards
  which are said to have infinite horizons.
  \begin{enumerate}
  \item 
    Denote by $\mM_{\textrm{free}} \subset \mM$ the subset of collision--free
    orbits, i.e.
    \begin{equation*}
      \mM_{\textrm{free}} = \{x \in \mM \,:\, S^{\bR}x
      \cap \partial \mM = \emptyset\}\,.
    \end{equation*}
  \item 
    The billiard has \emph{finite horizon} if $\mM_{\textrm{free}} =
    \emptyset$. Otherwise it has \emph{infinite horizon}.
  \end{enumerate}
\end{definition}
\noindent This notion applies to the models defined in Subsections
\ref{sec:Lproc}-\ref{ssec:wind} below. 

\subsection{Lorentz Process} 
\label{sec:Lproc}

The Lorentz process was introduced in
1905 by H. A. Lorentz \cite{L05} for the study of a dilute electron
gas in a metal.\footnote{In fact, Drude \cite{D00} had introduced a
  similar model as early as 1900. While the models of Drude and Lorentz 
  gave different ratios between the thermal and electrical conductivities,
  both were in accordance with the empirically observed
  Wiedemann--Franz law and, in that respect, provided decisive early
  contributions to the kinetic theory of gases \cite{Hoffmann2006}.}
While Lorentz considered the motion of a collection of independent
pointlike particles moving uniformly among immovable metalic ions
modeled by elastic spheres, we consider here the uniform motion of a
single pointlike particle in  a fixed array of spherical scatterers with
which it interacts via elastic collisions\footnote{More generally, the
  model can be extended to strictly convex scatterers rather than
  spherical ones.  
}; see, however, Subsection \ref{sec:Lgas} below for the extension of
the process to many particles.

Thus defined, the \textit{Lorentz process} is the billiard dynamics of
a point particle on a billiard table $\mQ =  \bR^d
\setminus \cup_{\alpha=1}^\infty \, O_{\alpha}$, where the scatterers
$O_{\alpha}$,  $ 1 \leq \alpha \leq  \infty$, are strictly convex with
$C^3$-smooth boundaries. Generally speaking, it could happen that
$\mQ$ has several connected components. For simplicity,
however, we assume that the scatterers are disjoint and that
$\mQ$ is unbounded and connected.  The phase space of this process is
then given according to the above definition, namely $\mM=\mQ\times
\bS_{d-1}$. 

It should finally be noted that, under this assumption, 
the Liouville measure $\rmd\mu= \rmd q\, \rmd v$, while invariant, is
infinite. If, however, there exists a regular lattice of rank $d$ for 
which we have that, for every point $z$ of this lattice, $\mQ + z =
\mQ$, then we say that the corresponding Lorentz process is
\textit{periodic}. In this case, the Liouville measure is finite (more
exactly, its factor with respect to the lattice is finite).

\subsection{Lorentz Gas} 
\label{sec:Lgas}

A closely related object is the \textit{Lorentz gas}, by which me mean
the  joint motion of a countable number of completely independent
Lorentz processes\footnote{In our nomenclature, we thus make a
  distinction between the dynamics of a single particle and that of
  countably many of them. The former case refers to the 
  \textit{process}, or a \textit{flow}, defined in the previous
  Subsection, and the latter to a \textit{gas}. This usage differs
  from that adopted by many authors who use 
  the notions of gas and process interchangeably.}. By 
keeping with our previous notation, the phase space of the Lorentz gas
is thus $\mM^\infty = \Pi_{j = 1}^\infty \mM_j$ where each $\mM_j$ is
a copy of $\mM$, the phase space of the Lorentz process. 
Furthermore, we only consider points in $\mM^\infty$
which are locally finite, that is, for every bounded $\mA \subset \bR^d$, 
$\sum_{j = 1}^\infty \mathbbm{1}_{q_j \in \mA} < \infty$. 

Now the smooth invariant measure of the dynamics is a Poissonian
measure\footnote{It is at the same time the Gibbsian measure.} in
$\mM^{\infty}$, with a uniform density; see \cite{MS19} for more
details. 

We note here that, in principle, we could permit the velocity space of the
Lorentz gas to be the whole of $\bR^d$ rather than $\bS_{d-1}$, as, in
fact, did Lorentz who considered Maxwellian distributions of
velocities. However, since the energies of the particles are
individually conserved, this would be a trivial generalisation.

\subsection{Wind--Tree models} 
\label{ssec:wind}

The wind--tree gas was initially proposed by P. and T. Ehrenfest in 1912,
\cite[Appendix to Section 5]{EE12} as a ``much simplified model''
aiming to understand ``what the position of the Stosszahlansatz is in 
the Maxwell-Boltzmann investigations.'' Since then, it has been
extended and generalised in many ways.

Generally speaking, the \textit{wind--tree process} (or \textit{flow})
is analogous to the Lorentz process. Its distinctive feature, however,
is that it is not dispersing and, in that sense, is a neutral version
of it. The main difference with the Lorentz process is indeed that 
the scatterers of the wind--tree process are parallepipeds (rhomboids,
cuboids, \dots),  usually parallelly positioned\footnote{The
  Ehrenfests' wind--tree model thus consists of square obstacles
  positioned at random on the two-dimensional plane, with their
  diagonals parallel to the plane's axes}. Consequently 
no hyperbolicity is present.  

By extension, the \textit{wind--tree gas} consists of countably many
independent copies of wind--tree processes. Notations analogous
to those of Subsection \ref{sec:Lproc}-\ref{sec:Lgas} carry over.

We first mention two planar models with identical rectangular
scatterers, whose sides are parallel to the coordinate axes.

\subsubsection{Aperiodic wind trees}
\label{sec:aperwindtree}

The first one of them, actually a family of models, was studied in 
\cite{M-ST16}. The authors consider the set of unit square cells with
\emph{square} scatterers of sides $2\,r$, $\tfrac{1}{4} \leq r < 
\tfrac{1}{2}$, centered at points $(a, b)  \in[0, 1]^2$, and such that
the scatterers are contained within the unit cells.  This set may thus
be parametrised by
\begin{equation*}
  \mA = \{(a, \, b): r \le a,\, b  \le 1-r\}\,,
\end{equation*} 
with the topology inherited from $\bR^{2}$. On the plane, the
parameter space is $\mA^{\bZ^2}$, with the product topology. Then each
parameter value $g = \{(a_{i, j}, b_{i, j}): 
{(i, j) \in \bZ^2}\} \in \mA^{\bZ^2}$ defines a
wind--tree billiard in the plane, with the collection of square
scatterers $O_{i, j}$, each centered at point $g_{i, j} +(i,\, j)$, $(i, \,j) 
\in \bZ^2$.  

In this case the billiard table is $\mQ_g = \bR^2 \setminus
\cup_{(i, j) \in \bZ^2} \, O_{i, j}$ and the dynamics 
is $S_g^t: \mM_g \to \mM_g: -\infty < t < \infty$, with phase space
$\mM_g$ defined in Section \ref{sec:discvel} below. This billiard is
called the wind--tree process. 

\subsubsection{Periodic wind trees}
\label{sec:perwindtree}

The second model is a periodic version of the wind--tree model,
introduced in \cite{Hardy1980} and investigated more recently in
\cite{FU14, DHL14}.  The scatterers are upright isomorphic
\emph{rectangles} 
$O_{i, j}$ with sides of lengths $0 < a, b < 1$ and centered at the lattice
points of $\bZ^2$. 

In this case, the billiard table is $\mQ_{a, b} = \bR^2 
\setminus \cup_{(i, j) \in \bZ^2}\, O_{i, j}$ and the dynamics on this table,
$S^{t}: \mM \to \mM: -\infty < t < \infty$, is a billiard, i.e.~ a
wind--tree process. We now turn to the definition of phase space
$\mM$. 

\subsubsection{Discrete velocity space}
\label{sec:discvel}

A characteristic of wind--tree models is, in general, that the set of
possible directions for the billiard flow is finite. The Ehrenfests
themselves thus considered the space of four velocity directions $(\pm
1,0)$, $(0,\pm1)$. 

If we allow for an arbitrary initial angle $\theta$, $0 < \theta <
2\,\pi$, the vertical and horizontal reflections on the plane generate
the set of four different possible directions $\{\pm \theta, \pm(\pi -
\theta)\}$ the particle can take at any time.  Letting $[{p}]$
denote the set of the corresponding unit velocity vectors, 
\begin{equation*}
  [{p}] = \Big\{v \in \bS_{1}: \arctan \frac{v_2}{v_1}  
  \in \{\pm \theta, \pm(\pi - \theta)\}\Big\}\,,
\end{equation*}
the phase space, in the case of the aperiodic wind--tree model, is then
$\mM_g = \mQ_g \times [{p}]$ and, in the case of the periodic
one, $\mM = \mQ_{a, b} \times [{p}]$. 

The corresponding invariant measures of the flows $S^{t}_{g}$ and $S^{t}$ are
in both cases the infinite measures $\rmd\mu \propto \rmd q$ (up to
the counting measure for the velocity space).

\subsubsection{Higher-dimensional cases}
\label{ssec:highwindtree}

In Subsection \ref{ssec:randwt} we will also recall results on a
\textit{random wind--tree model} in $\bR^3$. There the scatterers
form an array of randomly placed, identically oriented cubes with
sides again parallel to the coordinate axes. As with the planar
case, the velocities of the particle form a finite set,
this time with eight elements. Indeed, following \cite{LT19}, fix a
probability vector ${p} = (p_1, p_2, p_3)$ with $ p_i > 0\ \
\forall i$ and let $|{p}| = \sqrt{p_1^2 + p_2^2 
+ p_3^2}$ denote its norm. Then the set of possible velocities is
\begin{equation*}
  [{p}] = \Big\{v \in \bS_2: |v_i| =
  \frac{p_i}{|{p}|}\Big\}
  \,.
\end{equation*}

\subsection{Hard Ball Systems}
\label{ssec:hardballs}

Assume that, in general, a system of $N : N \ge 2$, identical (for
simplicity) balls of unit masses 
and radii $r>0$ are placed at non-overlapping positions in 
$\bT^\nu=\bR^\nu/\bZ^\nu$, the $\nu$-dimensional unit torus, 
$(\nu\ge 2)$, and given random velocities $v_{i} : 1 \leq i \leq N$.
The dynamics corresponds to the uniform motion of the ball particles
with elastic collisions when they get into contact.  

Denote the phase point of the $i$\textsuperscript{th} ball by
$(q_i,v_i)\in \bT^\nu\times \bR^\nu$. The 
configuration space $\mQ$ of the $N$ balls is a subset of
$\bT^{N\,\nu}$, obtained from $\bT^{N\,\nu}$ by cutting out
the $\left({N\atop 2}\right)$ cylindric scatterers,
\begin{equation*}
  C_{i,j} =\left\{ (q_1,\dots,q_N)\in \bT^{N\,\nu}:\mid q_i-q_j\mid
    < 2r\right\},
\end{equation*}
$1\leq i<j\leq N$. That is, $ \mQ:= \bT^{N\,\nu}\setminus
\cup_{1\le i < j \le N} \, C_{i,j}$.  The (kinetic) energy $K=\tfrac{1}{2}\,
\sum^N_1 m_i \, v_i^2$ and the total momentum $P=\sum^N_1 m_i \, v_i$
are first integrals of the motion. Thus, without loss of generality, we
can assume that $K=\tfrac{1}{2}$, $P=0$. (If $P\not= 0$, then the system
has an additional conditionally periodic or periodic motion.) Now, for
these values of $K$ and $P$, we define our dynamical system.

The set $\mQ$, a compact, flat Riemannian manifold with boundary, such
as identified above, is the configuration space of our
system. Its phase space is $\mM:= \mQ \times \bS_{N\,\nu -1}$.
The Liouville measure $\rmd \mu = \mathrm{const.} \, \rmd q \, \rmd v$
is invariant with respect to the evolution $S^{\bR} := \{S^{t}:\ t \in
\bR\}$ of our dynamical system defined by elastic collisions of the
balls of unit masses and their uniform free motion. The dynamics can,
indeed, be defined for $\mu-$a.~e. phase point. This is a billiard
which, after restricting to $P=0$ and fixing the center of mass,
is dispersing for $N = 2$ and semi-dispersing for all $N \ge 3$.

\subsection{Systems of spatially localised Hard Balls}
\label{ssec:spatial}

After the ergodicity of gases of two hard balls \cite{S70,BS73,SCh87},
then three \cite{KSSz91}, and four \cite{KSSz92}, had been
established, Bunimovich \emph{et al.} \cite{BLPS92} observed
that
\begin{quote} 
  Unfortunately, new and serious technical problems, which
  require the development of some specific methods, appear at each step
  from $N$ to $N + 1$ balls.
\end{quote} 
To go around this difficulty, Bunimovich \emph{et al.} \cite{BLPS92}
put forth a family of models that, as they write, ``are intermediate
ones between the gas of  hard balls and the Lorentz gas model'' and
thus lend themselves to a systematic study of ergodicity for arbitrary
number of particles. While these models are gases of hard balls (in
the usual sense of the expression), they exhibit the distinctive
feature that individual balls are trapped in their own cells and
thereby retain a form of spatial order. Models in this class thus
feature both the collisional dynamics of a gas and cristalline spatial
structure of a solid.   

As far as ergodicity is concerned, the fundamental advantage of these
models over other systems of hard balls is that the collision sequences
of the particles are much simpler and easier to control. Another
important feature is that, while the models allow for heat transport
through energy exchanges, they prevent mass transport. This raises the
prospect, as emphasised by the authors of \cite{BLPS92}, that
\begin{quote} 
  it would be interesting to investigate the kinetic
  properties of these models, e.~g.  diffusion of energy.
\end{quote}
While the \emph{raison d'\^etre} of the models, i.e.~ergodicity of a
gas of any number of particles, has been superseded by
the understanding of the inductive step alluded to above and 
the ultimate proof of the ergodic hypothesis for hard ball systems
\cite{S19}, the point above has been key to continued interest in 
these models and we will comeback to it in the sequel. 

\subsubsection{Bunimovich--Liverani--Pellegrinotti--Suhov models}
\label{ssec:BLPS}

Here we consider a specific planar version of the models, which easily lends
itself to various generalisations in two and higher spatial dimensions.  

\begin{figure}[hbt]
  \centering
  \includegraphics[width=.8\textwidth]{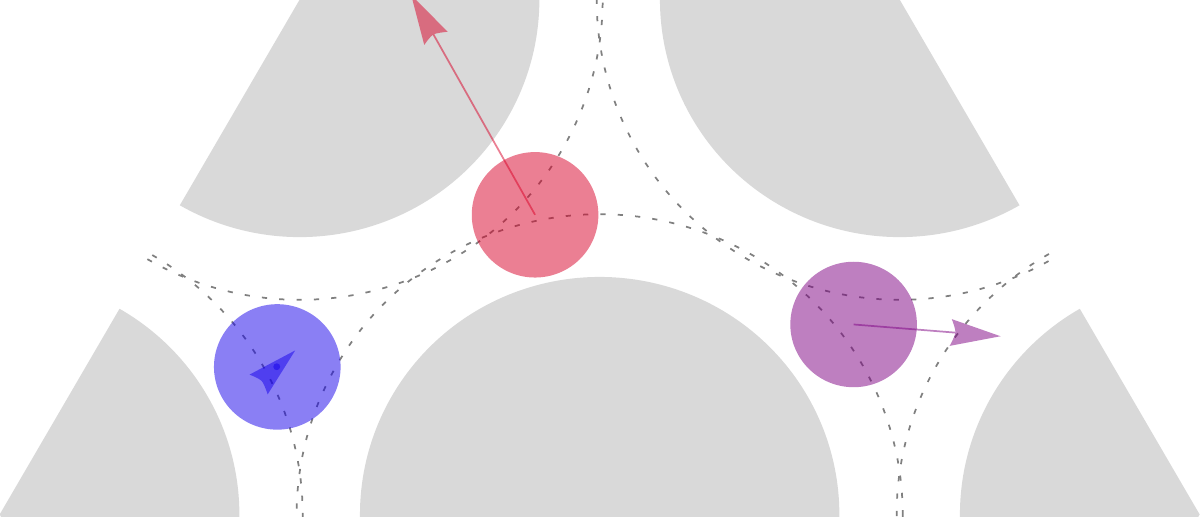}
  \caption{An illustration of the two-dimensional version of the
    BLPS model built on a honeycomb lattice. Fixed scatterers of radii
    $\rhof$ are coloured in gray and moving balls of radii
    $\rhom$ in different hues of blue and red. Differences
    in these colors, as well as in the velocity arrow sizes, reflect
    differences in the values of kinetic energy. The dotted circles
    have radii $\rhom + \rhof$. 
  } 
  \label{fig:blps}
\end{figure}

The model is illustrated in Figure \ref{fig:blps}
and can be constructed according to the following few 
steps. Consider a regular honeycomb lattice on the plane. Unit cells
alternate between upward and downward equilateral triangles of unit
sides. Around each lattice point, discs of radii
$\rhof$,  $0 < \rhof < \tfrac{1}{2}$, are
removed from the plane, which serve as fixed scatterers. 
Inside every triangle, let there be one circular moving ball of unit
mass and radius $\rhom$, $\tfrac{1}{2} - \rhof
< \rhom < \tfrac{1}{\sqrt{3}} - \rhof$. Their
positions must be chosen 
so that moving balls do not overlap with any of the fixed
scatterers, as well as among each other.

The lower bound $\rhof + \rhom > \tfrac{1}{2}$ ensures that each disc
remains in the cell it starts from; see overlap among dotted circles in
the figure. The upper bound $\rhof + \rhom < \tfrac{1}{\sqrt{3}}$ is so as
to leave enough room to fit the moving ball in the remaining space.

To allow for collisions among moving balls, however, we must  
further assume that the balls are large engouh, i.e.
\begin{equation*}
  \rhom > \rhoc = \sqrt{(\rhof + \rhom)^2 - \tfrac{1}{4}}
  \,.
\end{equation*}
Otherwise there would be no interaction and the model would be of
little interest. 

Given initial positions and velocities, the moving balls follow the
billiard dynamics. That is, the balls move uniformly until an elastic
collision event occurs, either among two moving balls, or against a
scatterer. 

The model introduced above is, of course, one with an infinite number
of particles. However, it lends itself to different restrictions with
a finite number of particles, such as a one-dimensional chain of
alternating upward and downward cells as illustrated by the figure,
or, in its simplest form, a gas of two moving balls, each trapped in
their own cells. The corresponding billiard in this minimal case is
a four-dimensional semi-dispersing one. 

\subsubsection{B\'alint--Gilbert--N\'andori--Sz\'asz--T\'oth model of
  pistons and balls}
\label{ssec:BGNST}

A variant of the BLPS model was proposed by B\'alint \textit{et al.}
\cite{BGNSzT17}. Unlike the former model, which, as described
above, consists of a collection of similar unit cells spanning the
two-dimensional plane, the latter model consists of a hybrid
juxtaposition of one-dimensional-like particles, called 
pistons, and two-dimensional ball particles. Energy exchanges in the
system are therefore mediated by ball-piston interactions.

\begin{figure}[hbt]
  \centering
  \includegraphics[width=.9\textwidth]{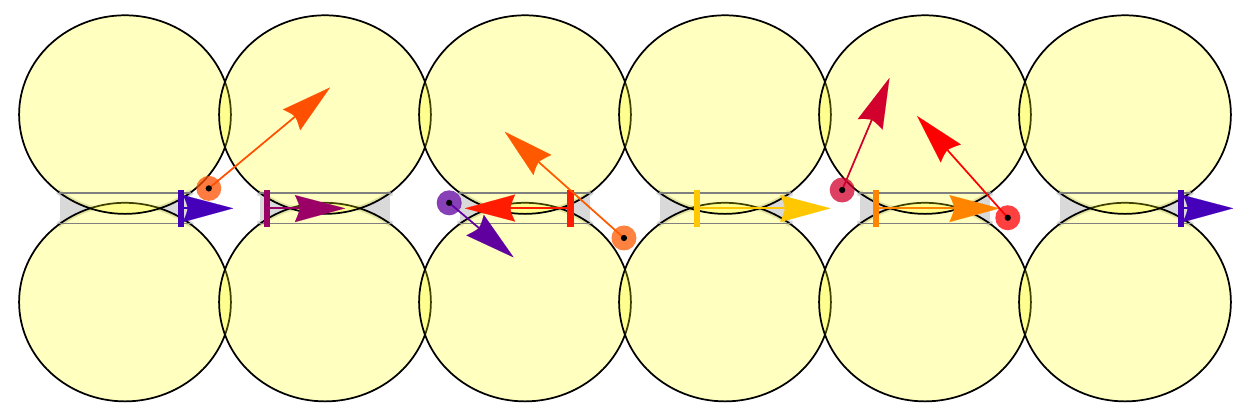}
  \caption{An illustration of the BGNST model alternating ball
    and piston cells along a one-dimensional lattice. In this specific
    example, periodic boundary conditions apply, so that the left-- and
    right--most pistons are identical. The color-coding and arrow
    sizes are as described in the caption of Figure \ref{fig:blps}.
  } 
  \label{fig:bgnst}
\end{figure}

Here we consider linear chains such as shown in Figure
\ref{fig:bgnst}.  While point particles (balls in figure) move about
their respective planar cells in the usual fashion of two-dimensional
billiards, the motion of pistons is restricted to horizontal line
segments (represented as rectangles in the figure), along which they
move back and forth. Importantly, the lengths of these cells are large
enough that piston particles, which have a fixed vertical span,
penetrate into the two neighbouring ball cells and may thus
mediate energy exchanges between neighbouring balls through successive
collisions. The system so-defined is a ball-piston gas and obeys the 
billiard dynamics.

An interesting property is that the minimal model consisting of a
single pair of ball and piston can be viewed as a point particle
moving about a three-dimensional cavity of cylindrical shape with a
corner cut out, i.e.~a three-dimensional semi-dispersing billiard; see
\cite{BGNSzT17} for further details. As far as controlling the decay
of correlations between particles in neighbouring cells, this renders
the BGNST model more easily tractable than the BLPS model. 

\subsection{Rayleigh Gas}
\label{ssec:rayleigh}

In 1891 Lord Rayleigh \cite{Rayleigh1891} introduced a model system of
binary mixture to study relaxation to equilibrium in the framework of
the theory of gases. Originally
posed as a one-dimensional model, this is a gas of two types of hard
balls, one very small  relatively to the other. It was in fact the first
theoretical attempt towards the description of Brownian motion. As one
of its limiting cases, it contains the Lorentz gas.

For our purpose, the \textit{Rayleigh gas} is a binary gas of hard
ball particles in $\bR^d: d \ge 1$, whereby a single particle, called
the \textit{Rayleigh particle}, with mass $M >0$ and radius $r > 0$,
interacts with a collection of point particles of unit masses
oblivious to one another. 
Having specified the positions and velocities of all particles, the
gas thus evolves according to billiard dynamics, i.e.~uniform 
motion until an elastic collision occurs when the Rayleigh particle and one
of the point particles come into contact;  see \cite{S91,SzT87} for
further details.

So defined, the dynamics has, of course, no invariant probability
measure. Several modifications have thus been considered, some of
these are described below.

\subsubsection{Coordinate system fixed to the Rayleigh particle}
\label{ssec:rayleighfixed}

The dynamics with coordinate system fixed to the Rayleigh particle has
been termed the \textit{M\"unchhausen picture}
\cite{SzT87}\footnote{It is worth mentioning that the work of
  reference \cite{SzT87} was  motivated by Sinai's Tur\' an Memorial Lectures
  held in Budapest in 1985; see \cite{SS86, SzT86} }. 
This system admits an invariant probability measure in the form of 
\begin{enumerate}
\item 
  the product of a Poissonian measure for the gas of point particles in
  the product space of positions outside the Rayleigh particle
\item
  with Gaussian distribution of their velocities as well as that 
  of the Rayleigh particle (with variance depending on its mass parameter $M$).
\end{enumerate}

\subsubsection{Semi-permeable Rayleigh gas} 
\label{ssec:rayleighsemi}

An alternative is to restrict the motion of the center of the Rayleigh
particle to a finite region---say a sphere of radius $R$---acting as
permeable for the point particles and reflecting for the Rayleigh
particle. This dynamics has a natural invariant probability measure
which is the product of 
\begin{enumerate}
\item a uniform spatial distribution for the center of the Rayleigh
  particle in the sphere of radius $R$,
\item
  under the condition that the center of the Rayleigh
  particle is selected, the atoms are distributed according to a Poisson
  point distribution in the complement of the $R$-sphere of the Rayleigh
  particle, and
\item
  as above, a Gaussian distribution of the velocities.
\end{enumerate}

\subsubsection{Rayleigh gas on the half line} 
\label{ssec:rayleighline}

On the positive half-line the distance of the Rayleigh particle has a
probability distribution therefore the whole system has a
time-invariant distribution. In fact, denote the positions of the
particles by $0 < X < x_1 < x_2 < \dots$ where $X$ is the
position of the Rayleigh particle and $(x_1, x_2, \dots)$ are the
positions of the point particles. The invariant probability measure is
the product of 
\begin{enumerate}
\item  
  a Poisson process on $\bR_+$ and
\item  
  as above, a Gaussian distribution of the velocities.
\end{enumerate}

\section{Boltzmann's Ergodic Hypothesis}
\label{sec:ergo}

At its core, statistical mechanics aims at characterising the
properties of matter in the bulk by accounting for the contributions
of its constitutive molecules. From the point of view of dynamical
systems, a central problem is to understand the emerging behaviour of
systems whose states evolve according to a set of ordinary or
partial differential equations, with an emphasis on Hamiltonian
systems, or, alternatively, whose evolution is specified by an
iterative system, with emphasis on symplectic maps.

The 1960's and 1970's saw spectacular progress in providing a 
mathematically rigorous basis for statistical mechanics. 
In particular, a new and rich arsenal of mathematical methods was
discovered, which includes, to mention but a few examples,
Kolmogorov--Arnold--Moser theory, renormalisation group methods,
bifurcation theory, the theory of strange attractors, the hyperbolic
theory of dynamical systems, thermodynamic formalism, Lyapunov
exponents and entropy theory, Ornstein's theory of
isomorphisms\footnote{The volume \cite{VW85}, based on a 1983 school
  devoted to regular and chaotic motions in dynamical systems,
  summarises some of the main achievements at the time, with, in
  particular, a substantive and rich introductory paper by A. S. Wightman,
  \cite[\emph{Introduction to the Problems}]{VW85}, putting them in a
  historical perspective and thus providing an excellent snapshot on the
  early history of dynamical systems, starting from Newton, via Poincar\'e
  and Birkhoff. 
}.

Sinai's 1970 proof of the ergodicity of the gas of two hard discs on
the two-dimensional torus and his 1963 hypothesis on the ergodicity of
any number of hard ball particles, nowadays referred to as the Boltzmann--Sinai
ergodic hypothesis \cite{Sz96, S19}, was a substantial contribution to the
aforementioned list of notable breakthroughs. 

It must, however, be said that, in the list of stochastic properties
of a system,  ergodicity belongs to the weaker ones. There are 
several stronger such properties which we do not treat here; for
instance, mixing, K-mixing, or Bernoulli. Nevertheless, in Section
\ref{sec:diff}, we address another important qualitative property, that of
\textit{correlation decay} and its rate, which is most significant for
the sake of assessing the diffusive properties of physical systems.

We recall below Boltzmann's ergodic hypothesis, Subsection
\ref{ssec:boltz}, which deals with the limit of particle number
increasing to infinity. Unfortunately, our understanding of this case
remains very limited, to the point that formulating the appropriate
statement remains a delicate issue. Before we attempt at this goal, we
therefore treat separately and in more details the two cases 
of fixed and finite number of particles, Subsection \ref{ssec:fixedN},
or infinite, Subsection \ref{ssec:infty}. We then allow ourselves to
provide some thoughts on the former case in Subsection
\ref{ssec:increase}. 

\subsection{Boltzmann's Ergodic Hypothesis}
\label{ssec:boltz}

The hypothesis is essentially contained in the following approximate
\begin{statement}[Boltzmann's Ergodic Hypothesis]
  For large systems of interacting particles in equilibrium,
  time averages are close to ensemble 
  averages.
\end{statement}

Before delving into the specifics of the ergodic hypothesis, we open a
parenthesis to make mention of its controversial implications 
for statistical physics. For instance, in discussing relaxation to
equilibrium, M. Kardar \cite[pp~61--62]{K07} writes~:
\begin{quote} 
This brings us to the problem of ergodicity, which is whether it is
justified to replace time averages with ensemble averages. In
measuring the properties of any system, we deal with only one
representative of the equilibrium ensemble. However, most macroscopic
properties do not have instantaneous values and require some form of
averaging. For example, the pressure $P$ exerted by a gas results from
the impact of particles on the walls of the container. The number and
momenta of these particles varies at different times and different
locations. The measured pressure reflects an average over many
characteristic microscopic times. If over this time scale the
representative point of the system moves around and uniformly samples
the accessible points in phase space, we may replace the time average
with the ensemble average. 
\end{quote}
One might infer from these words that the ergodic hypothesis is indeed
fundamental for the definition of such an elementary notion as 
pressure. Yet Kardar goes on to write~:
\begin{quote}
For a few systems it is possible to prove an ergodic theorem, which
states that the representative point comes arbitrarily close to all
accessible points in phase space after a sufficiently long
time. However, the proof usually works for time intervals that grow
exponentially with the number of particles $N$, and thus exceed by far
any reasonable time scale over which the pressure of a gas is
typically measured. As such the proofs of the ergodic theorem have so
far little to do with the reality of macroscopic equilibrium. 
\end{quote} 
The latter sentence may be interpreted to say that the relevant time
scale is that which controls the decay of correlations at the scale of
molecular motion (Kardar's characteristic microscopic time). Apart
from fluctuations which are essentially controlled by the underlying
number of particles, physicists expect measurements of macroscopic
observables such as pressure to yield consistent values provided their
timescale is large enough with respect to the correlations of
molecular motion \cite{Toda1992}. 

\begin{remark}[Equilibrium averages] 
  In what follows, equilibrium averages always refer to the
  \textit{microcanonical ensemble}, i.e.~averages with respect to the
  Liouville equilibrium  measure $\mu$ on the submanifold of the phase
  space specified by the trivial invariants of motion.  More
  precisely, the ergodic hypothesis states that if $f$ is an
  observable (i.e.~a bounded measurable function on the phase space
  $\mM$ of the system), then, as the size of the system (say the
  number $N$ of particles) and observation time $T$ both tend to infinity,
  \begin{equation*}
    \frac{1}{T} \int_0^T f(S^{t}(x))\,\rmd t \to \int_{\mM} f(x) \rmd\mu(x)
  \end{equation*}
  where $S^{t}(x)$ is the time evolution of the phase point $x \in
  \mM$.
\end{remark}

Boltzmann formulated his celebrated hypothesis in the sense reflected
by the above remark, for a gas with an increasing number $N$ of
particles. In his subtle hypothesis neither the mathematical sense of
the limits $N \to \infty$ or $T \to \infty$ nor their order were
precisely given and it did not satisfy the demands of accuracy
required by physics, much less by mathematics. While the ergodic
theorems\footnote{Many historical aspects of ergodic theory were
  recently reviewed in \cite{Moore2015} to commemorate these two
  ergodic   theorems on the occasion of PNAS 100th Anniversary; see
  also the accompanying commentary \cite{Ashley2015}.} 
of von Neumann \cite{N32} and of Birkhoff \cite{B31,BK32} were
instrumental mathematical achievements, it still remained completely
open whether systems of interest to physics are ergodic or not.  

In this sense, \textit{Sinai's precise formulation of the
  ergodic hypothesis} for a physical system with a fixed number of
degrees of freedom \cite{S63}---later called the
\textit{Boltzmann--Sinai ergodic hypothesis} \cite{Sz96, S19}---and,
moreover, his \textit{proof of the ergodicity of two colliding discs on
the two-dimensional torus} caused a sensation. The sequence of events
that took place between the works of Birkhoff and von  Neumann
\cite{B31, N32}, on the one hand, and Sinai \cite{S63,S70}, on the
other hand, are related in \cite{Sz96} and we refer to it for further
details.

We recall below the definition of ergodicity and otherwise refer to
the surveys \cite{S19,Sz96} or the monograph \cite{CFS82}.

\begin{definition} 
Let $(\mM, \, S^{\bR},\, \mu)$ (resp. $(\mM, \, T^{\bZ},
\,\mu$)) be a group of probability preserving maps.  In the case of a
flow (resp. map), a subset $A \subset \mM$ is called {\rm invariant}
${(\rm mod}\, 0)$ if for each $t \in \bR$ one has $\mu(S^{t}A\,
\Delta\, A) = 0$ for each $t \in \bR$  (resp. $\mu(A \, \Delta\,
T^{-1}A) = 0$). A flow (or a map) is {\rm ergodic} if all invariant 
subsets are trivial, i.e.~they have measure $0$ or $1$.
\end{definition}

If, in general, we are given a dynamical system defined by a
Hamiltonian $H$ of $N \ge 2$ particles on $\bT^\nu: \nu \ge 1$,
then by fixing its invariants of motion, the time-evolution of the
system provides a probability-preserving flow and the problem of
ergodicity is actually about the ergodicity of \textit{this flow}. In the
present section we fix the Hamiltonian (and thus the dimension, too)
and, as said above, we will separately---and in different
depths---discuss in Subsections
\ref{ssec:fixedN}--\ref{ssec:increase} the cases 
\begin{enumerate}
\item $N < \infty$,
\item $N = \infty$,
\item $N \to \infty$.
\end{enumerate}

\subsection{Fixed number $N$ of hard balls}
\label{ssec:fixedN}

This case was amply covered in the reviews \cite{S19,Sz96} and our
exposition will therefore be concise. The review \cite{B19} includes a
discussion about stadium and related billiards, which we do not treat
here. 

In words, one has in mind a system of $N$ particles on
$\bT^\nu: \nu \ge 1$ interacting via a smooth Hamiltonian $H$
and moving on the submanifold of the phase space specified by the
invariants of motion. Early results by Markus and Meyer \cite{MM74}
showed that in the space of smooth Hamiltonians both nonergodic
systems and ergodic ones form dense open subsets\footnote{Of course, these
subsets can both be very small so it may---and in fact does---occur
that none of them represents a `typical' behaviour.}.

\subsubsection{Boltzmann--Sinai Ergodic Hypothesis} 
\label{ssec:BSergodic}

Consider the dynamical system of $N$ identical, sufficiently small,
elastic hard balls moving on $\bT^\nu$. Take the submanifold of the
phase space specified by fixing its invariants of motion (energy and
momentum). Assume that the conserved  momentum vector is $0$, which
also allows to fix the position of the center of mass. This 
submanifold has dimension $d = 2(N - 1)\nu -1$ (deduction of $1$ for
the energy and twice $\nu$ for the momentum and center of mass).

\begin{statement}[Boltzmann--Sinai Ergodic Hypothesis]  
  The system of $N : N \ge 2$ elastic hard balls on $\bT^\nu : \nu \ge 2$
  is, for sufficiently small radius of the balls, ergodic on the
  submanifold of the phase space specified by the invariants of
  motion.
\end{statement}

\begin{remark}
  Simple arguments lead to the following corollaries:
  \begin{enumerate}
  \item If the hypothesis is true and the total momentum is not zero,
    then the motion is the product of an ergodic flow and a conditionally
    periodic motion;
  \item If the hypothesis is true, then there is a finite number of
    ergodic components in the case when the radius of the balls is not
    small.
  \end{enumerate}
\end{remark}

Sinai's highly acclaimed 1970 paper \cite{S70} on the ergodicity (and
even K-property) of what became known as two-dimensional
Sinai billiards with finite horizon drew on important prior works. On
the one hand, N. S. Krylov \cite{K50}, the great Russian theoretical
physicist, observed in 1942 that the interaction of hard balls
is hyperbolic in a sense similar to that which had appeared earlier in
the proofs of Hedlund \cite{He39} and Hopf \cite{Ho39} of the
ergodicity of geodesic motion in hyperbolic geometry. On the other
hand, those were exactly the results of Hedlund and Hopf which
motivated Anosov and Sinai to create a beautiful and 
far-reaching theory for smooth uniformly hyperbolic dynamical systems;
see \cite{S60,A67,AS67}. 

Sinai's 1970 work on two-dimensional billiards \cite{S70} 
was innovative in several
respects. As a brief side-note, it must be said that Sinai's 1970
paper is also a gem of mathematical style. Because of the abundance of
new and original ideas, Sinai had to find an appropriate balance between
completeness and conciseness, providing  sufficient amount of
information for readers to follow his arguments while keeping the
paper brief enough to be at all readable. Even so, his work was hard
to understand and, apart from results by the Moscow school, for many
years, developments of his theory were unsurprisingly few\footnote{In
  1974 Gallavotti \cite{G75} provided his own version of 
  Sinai's proof, whereas Gallavotti and Ornstein \cite{GO74} were able
  to use part of his results to go from the K-property of
  Sinai billiards to their Bernoulli property. We may
  further note that: (i) Gerhard Keller, in his MSc thesis \cite{K77},
  written under the guidance of Konrad Jacobs in Erlangen, 
  reproduced Sinai's original proof; (ii) Vetier was among the first
  who could delve deep into Sinai's method, applying it to a new
  model: the Sinai billiard in a potential field; see
  \cite{V84,V89}.}. 

After a long history consisting of several breakthroughs, which is
nicely recalled in \cite{S19}, Sim\'anyi \cite{S13} eventually
completed the proof of the Boltzmann--Sinai hypothesis for an
arbitrary number of balls in any dimension\footnote{Sim\'anyi's result
  also covers the case of arbitrary masses of the balls.}.

\subsection{Infinite number of particles}
\label{ssec:infty}

As said earlier, the $N \to \infty$ case---and the appropriate
formulation of the ergodic hypothesis---is hard. The $N = \infty$ case
is more easily amenable to study. Some of the interesting results are
surveyed below and some conjectures are drawn. 

Key features of the associated models we wish to emphasise are, on the
one hand,  the good spatial mixing property of the invariant (Gibbs)
measure, and, on the other hand, the fact that such spatial
mixing does indeed lead to nice ergodic properties.
In Reference \cite{GLA75}, this phenomenon is called \textit{escape of
  local information to infinity}. It appears in all results we
report below. 

We also remark that, in some instances, spatial mixing may coexist
with local mixing due to the interaction (more concretely its
hyperbolicity). While some proofs make use of the latter, we expect
the former will play a more prevailing role. 

\subsubsection{The ideal gas in $\bR^d: d \ge 1$}  

Here the invariant measure is the Poissonian measure in the product
space $\bR^d \times B$ where $B \subset \bR^d $ is arbitrary. The
proof of ergodicity is due to Sinai and Volkovysskii \cite{VS71}.

\subsubsection{The Lorentz gas in $\bR^2$} 

Sinai \cite{S79} showed that, under general conditions for an
otherwise arbitrary configuration of identical and fixed circular
scatterers, the planar Lorentz gas is ergodic. The proof exploits the
hyperbolicity of planar dispersing billiards. We expect that the
result can also be generalised to higher dimensions and therefore
propose the 
\begin{conjecture} 
  The \textit{Lorentz gas} in $\bR^d: d \ge 2$
  with spherical scatterers is ergodic under general conditions for an
  otherwise arbitrary configuration of identical and fixed scatterers.
\end{conjecture} 
Here it would be much interesting to understand
what is the minimal information that ensures ergodicity.  

\begin{remark} 
  Proving the ergodicity of multidimensional dispersing
  billiards is based on the so called theorem on local ergodicity.
  In particular, this theorem makes the assumption that scatterers are 
  algebraic \cite{BNSzT02}, which is expected to be only a technical 
  assumption. Nevertheless---beyond the self-sufficient appeal of the 
  Conjecture---it would be interesting to prove it for a class of 
  scatterers wider than merely algebraic ones.
\end{remark}

\subsubsection{Rayleigh gas} 

As explained in Subsection \ref{ssec:rayleigh}, one among the ways to
obtain a dynamical system with an invariant probability is to consider a
semi-permeable wall such that the Rayleigh particle be reflected
by this barrier while the gas particles go through it unscathed. The
ergodicity of the semi-permeable Rayleigh gas was shown for the case
$d = 1$ in \cite{GLR82} and generalised to $d \ge 2$ in \cite{ET90}.

The alternative half-line version of the Rayleigh gas ($d = 1$)
was also proven to be ergodic in \cite{BPPSS85}.

The Rayleigh gas can be understood as an infinite-dimensional
billiard where collisions with the Rayleigh particle correspond to
collisions with a cylindrical scatterer. A very weak form of
hyperbolicity is therefore present, but the aforementioned results do
not use it explicitly.

\subsubsection{Wind tree gas}
\label{ssec:windconjectures}

The theory of wind--tree processes has recently been the subject of
spectacular new and important results, partially thanks to the results
on the Lyapunov exponents of Kontsevich-Zorich cocycles
\cite{Zorich2006}.  
In the Appendix, we mention two different, relatively
simple results of this theory pertaining to the models discussed in
paragraphs \ref{sec:aperwindtree} and \ref{sec:perwindtree}, which
allow us to formulate below two 
conjectures about the ergodicity of the corresponding wind--tree
gases. Our goal is to emphasise that the wind--tree gas could be
ergodic without need for local hyperbolicty of the dynamics!

\begin{conjecture} 
  The aperiodic Ehrenfest wind--tree gas whose
  scatterer configuration satisfies the conditions of the wind--tree flow
  of \cite{M-ST16} is ergodic (cf. Theorem \ref{thm:M-ST} of
  Appendix \ref{sec:Ahom}
).
\end{conjecture}

\begin{conjecture}
  \label{conj:WTG} 
  Under the conditions of
  \cite{DHL14} on the shape parameter of the rectangles, the periodic
  wind--tree gas is ergodic (cf. Theorem \ref{thm:DHL} of Appendix
  \ref{sec:Ahom}). 
\end{conjecture}

\begin{remark}
  Let us emphasize again the difference between the wind--tree
  \emph{process} (which describes a single particle) and the wind--tree
  \emph{gas} (which describes infinitely many independent
  particles). In particular, by the results of reference~\cite{FU14},
  even after restriction to the subset of Subsection~\ref{sec:discvel},
  the periodic wind--tree process is not ergodic for almost all
  parameters.  
\end{remark}

\subsection{Number of particles increasing to infinity}
\label{ssec:increase}

As mentioned before we know very little about this most fundamental
problem whose goal would be to clarify the situation around
Boltzmann's ergodic hypothesis. Its importance to both quantum and
classical realms is well summarised by Penrose and Lebowitz
\cite{Lebowitz1973}: 
\begin{quote}
  This lack of knowledge is regrettable because only for an infinite
  system (by which term we mean the limit of a finite system as its
  size becomes infinite) can one expect to find strictly irreversible
  behavior in quantum mechanics. Moreover, the distinction between
  microscopic and macroscopic observables, which appears essential to
  any complete theory of irreversibility and kinetic equations, can
  only be formulated precisely for infinite systems.
\end{quote}

We refer to the following two quotes. The first 
one is due to Wightman \cite[\emph{Introduction to the Problems},
p~20]{VW85}: 

\begin{quote} There are three traditional reasons often given for the
legitimacy of the traditional methods, despite non--ergo\-di\-ci\-ty
of the flow.
  \begin{enumerate}[label=\Alph*]
  \item (Thermodynamic limit) The results of classical statistical
mechanics may be valid in the \emph{thermodynamic} or
\emph{bulk limit} (number of degrees of freedom, $N \to \infty$;
volume, $V \to \infty$; $\frac{N}{V} \to \rho < \infty$). This might
happen because the relative phase volume of the non-ergodic portion of
the flow approaches zero in the thermodynamic limit, so that
observables become insensitive to the non-ergodic portion.
  \item (Macroscopic observables) There might be a res\-tric\-ted
    class of observables, called \emph{macroscopic}, to which classical
statistical mechanics would apply in the thermodynamic limit. In that
limit, the macroscopic observables might be insensitive to the
non-ergodic portion of the flow even if its relative phase volume does
not go to zero.
  \item (Grain of Dust or Heat Bath) The idealization that the system
under study is isolated should be made more realistic by the inclusion of
coupling to outside systems. Then the extreme sensitivity of the
non-ergodicity to initial conditions might result in an averaged
behavior consistent with classical statistical mechanics.
  \end{enumerate}
\end{quote} 

The second quote we wish to refer to is due to Dobrushin \cite{D94} who,
assessing the nonfullfilment of the ergodic hypothesis with respect to
the foundations of statistical mechanics, offers that:

\begin{quote} 
For example, it is possible that even the Ergodic Hypothesis is
not valid and there are several ergodic components and, for large $N$
one of these components covers the main part of the phase
space. Another variant seems more plausible. For large $N$ there is a
lot of small ergodic components which are mixed in a so complex way
that using an observation in a fixed volume we almost can not
distinguish between these components. It is difficult to formulate
exactly such hypothesis and even more difficult to deduce its
implications.
\end{quote} 

The reader will appreciate that Dobrushin's first hypothesis
corresponds to Wightman's reason (A), and the second one 
to reason (B). There is unfortunately little progress to report on
these questions.

\section{Brownian motion}
\label{sec:diff}

One of the fundamental cornerstones of contemporary science, the
\textit{atomic theory of matter},  did not win universal
acceptance until the early 20\textsuperscript{th} century. While
a form of the atomic theory was taught as early the
4\textsuperscript{th} and 5\textsuperscript{th} centuries~BC by
Democritus \cite{Democritus}, its ultimate approval was the result of
\begin{enumerate}
\item 
  Einstein's 1905 \cite{E05} ingenious characterisation of Avogadro's
  number through his derivation of the diffusion equation on the basis
  of atomic theory,
\item 
  followed by Perrin's experimental determination \cite{Perrin1909} of
  Avogadro's number based on Einstein's arguments, which later
  ear\-ned Perrin the 1926 Nobel Prize in physics ``for his work on
  the discontinuous structure of matter, and especially for his
  discovery of sedimentation 
  equilibrium.''  
\end{enumerate} 
It thus naturally became a  paramount objective within the
mathematical physics community to derive Brownian motion ab initio,
i.e.~starting from the framework of Hamiltonian mechanics. Many
surveys of Einstein's work on Brownian motion and the many results it
inspired are available in the literature. Here we limit our references
to Duplantier's centenary survey \cite{D06} for general background and
the two recent reviews \cite{Sz14,Sz19},  which cover the approach to
diffusion via billiard theory. 

The emergence of Brownian motion for the Lorentz process relies on the
decay of correlations. The most widely used approach in ergodic theory is via
appropriate upper bounds on the correlation decay of some nice 
functions. In Subsection \ref{ssec:random} below, we begin with some
early results, which are rather based on random walks and where
correlations are actually absent. We discuss some of their
implications for limiting regimes of billiard dynamics which 
are widely used in the physics literature. As far as the mathematical
theory of billiards is concerned, our treatment of two-dimensional
models in Subsection \ref{ssec:Markov} will be concise, as details are
available in the aforementioned two recent reviews. The interesting
and still open case of dimension $d \ge 3$ is the topic of Subsection
\ref{ssec:multidim}. Finally, in Subsection \ref{ssec:randwt}, recent
results applying to the random Lorentz and wind--tree processes are
reviewed.

\subsection{Random walks: absence of correlations}
\label{ssec:random}

The simplest mathematical formalisation of Brownian motion is given by
the Wiener process \cite{Wiener1923}. An elementary mathematical
result for its derivation from underlying probabilistic laws was the
classic claim of reference~\cite{EK46} (see also \cite{D51,P56}), namely that
the Wiener process arises as the diffusive limit of a simple symmetric
random walk (with discrete time step); see~\cite{Kac1947}.
In some sense, the underlying probabilistic dynamics
is here specified at a mesoscopic level of description, finer than the
Wiener process, which defines the macroscopic evolution, but coarser
than a hypothetical deterministic law at the microscopic scale.  

In that respect, it is interesting to note that the baker's
transformation\footnote{While Hopf does not make use of the name
  ``baker's map'' in his monograph, he does remark that ``the repeated
  execution of which is reminiscent of the production of puff
  pastry.''} had been introduced earlier by Hopf
\cite[Paragraph 12]{Hopf1937}, who established the isometry with 
the Bernoulli process $B(\tfrac{1}{2},\,\tfrac{1}{2})$; see also \cite[Appendice
7]{Arnold1967}. The baker's transformation is both 
area-preserving and time-reversible, which confers it a special status
as it may be conceived of as a caricature of a Hamiltonian system
whose statistical properties are straightforward. While physicists
focused mostly on its mixing properties and were careful to warn
against  giving it too much significance \cite{Lebowitz1973}, it is
interesting to note that the baker's transformation composed with
translations on the lattice provides a simple and straightforward
deterministic law which can be interpreted as a microscopic
representation of the Wiener process \cite{Gaspard1992}.

\begin{figure}[hbt]
  \centering
  \includegraphics[width=.7\textwidth]{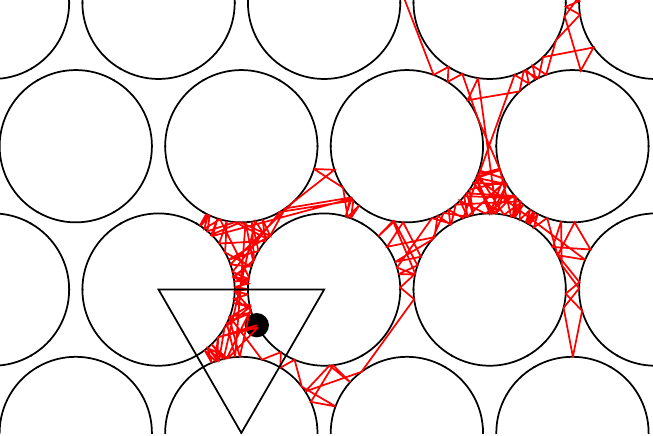}
  \caption{Diffusion in a finite-horizon Sinai billiard may be
    described by a random walk in the limit of vanishing spacing
    between obstacles. In physics literature, this is referred to as
    the Machta-Zwanzig regime \cite{MZ1983}.
  } 
  \label{fig:2dlg}
\end{figure}

In the case of the planar periodic Lorentz process with finite
horizon, such as depicted in Figure \ref{fig:2dlg}, the picture is
obviously complex. Suffice it to say, however, that convergence to the
Wiener process follows essentially from weak bounds on correlation 
decay  (these bounds are required to be at least summable); see
Subsection \ref{ssec:Markov}. 

A regime of particular interest is that of narrow spacing between
circular obstacles, which, in the physics literature, is often
referred to after Machta and Zwanzig \cite{MZ1983}. If one considers
the motion of a tracer particle between the triangular cells of the
honeycomb lattice of Figure \ref{fig:2dlg} and increases the diameter of
the obstacles to near the lattice parameter value, the hopping events
of the tracer particle between neighbouring cells become 
asymptotically rare, so that correlations between two such successive
events are virtually absent. The distribution of hopping times,
i.~e. the return times to the boundary segments separating
neighbouring cells, is then expected to follow an exponential law
\cite{Hirata1999, Pene2010}. At the level of lattice cells, the
tracer's dynamics thus reduces, to a continuous-time Markov jump
process. Moreover, and much like its discrete-time counterpart
referred to above, this process has a straightforward diffusive
scaling limit, which yields a  Wiener process with diffusion
coefficient given according to a dimensional estimate, that is, one
fourth the lattice constant squared multiplied by the properly scaled
average hopping rate. Furthermore, similar results can be derived in
arbitrary dimensions (under the finite horizon condition and so long
as correlations can be ignored---a very strong claim!).

\subsection{Markov techniques for Sinai billiards}
\label{ssec:Markov}

As said earlier, we limit our discussion to list the main techniques
used in the study of \textit{finite horizon Sinai
  billiards}\footnote{We refer to \cite{Sz19} and references therein for a
  review of infinite horizon Sinai billiards in the plane, and to
  \cite{D12, N14} for the multidimensional case.}. They are:
\begin{enumerate}
\item 
  \textit{Markov partitions} and stretched exponential correlation
  decay,
\item 
  \textit{Markov sieves} and stretched exponential correlation decay,
\item 
  \textit{Young towers} and exponential correlation decay,
\item
  \textit{Standard pairs};
\end{enumerate} 
see references~\cite{BS80,BS81,BS86} for Markov partitions,
\cite{BChS90,BChS91} for Markov sieves, \cite{Y98} for Young towers,
and  \cite{ChD07} for standard pairs.
The latter is a kind of approximate Markov partition whose great
advantage is that it also lends itself to the description of a family of
singular, hyperbolic dynamics obtained from one another by continuous
perturbations.

Markov techniques are primarily used as tools to handle
the correlation decay of physically or geometrically significant
functions. One thus obtains probabilistic limit theorems for these
functions that are central to the analysis of the relevant transport
processes. The case of Brownian motion is discussed in some detail
in the review article \cite{Sz19}.
It is also interesting to note that,  as an application to the momentum
current, the same techniques were exploited in \cite{BS96} to
establish the existence of both shear and bulk viscosities of the
solid phase of a periodic system of two hard balls on the
two-dimensional torus, a model completely analogous to the Sinai
billiard. We further refer to \cite{V03} where the extension of these
results to the liquid phase on a hexagonal torus was
investigated\footnote{
  In contrast to the square geometry where the liquid phase is
  associated with an infinite horizon regime, the hexagonal geometry
  exhibits a liquid phase which spans both finite and infinite horizon
  regimes.
}. 

Before going into further details about billiards, we make the following
\begin{remark} 
  In probability theory there is a wealth of results about the most
  delicate and detailed properties of convergence to Wiener process of a
  random walk (or on the closeness of these two processes in various
  senses).  Nowadays---in parallel with the aforementioned progress on
  the theory of Sinai billiards---one can observe a quite similar, rich
  and much promising development. Namely, what had earlier 
  been known for sums of independent variables is becoming a goal to
  be ascertained for the Lorentz process, i.e.~for sums of variables
  determined by a deterministic motion.
  Here we refer to \cite{Dolgopyat2008, Dolgopyat2009, Pene2009,
    Marklof2014, Pene2019, AL18}.
\end{remark}

It is our opinion that the aforementioned advances have been
strongly consolidated in the last decades by the publication of the
monograph by Chernov and Markarian \cite{ChM06}, which has
tremendously helped in promoting research in this highly technical
subject. Similarly, interest in the alternative functional analytic
method of \textit{transfer operators} greatly benefited from the
monograph by Baladi \cite{B00}. A recent and notable breakthrough in
the application of this method is the result of Baladi, Demers and
Liverani \cite{BDL18}, establishing exponential decay of correlations
for two-dimensional finite-horizon Sinai billiard flows without corner
points.

\subsection{The complexity hypothesis for higher-dimensional
  ($d \ge 3$) billiards}
\label{ssec:multidim}

The contents of the previous subsection are unfortunately
restricted to the planar case. It would be utmost important to
understand the multidimensional situation, which is much different
from the planar one.

As to \textit{correlation decay} for multidimensional billiards, we
know of only one result, due to B\'alint and T\'oth
\cite{BT08}. Assuming finite horizon and smooth boundary pieces, it
makes the following 
\begin{statement}
  Under the complexity hypothesis\footnote{After Sinai's
    original work \cite{S70}, the complexity hypothesis has been used
    in every work dealing with hyperbolic two-dimensional
    billiards. Roughly speaking it says that hyperbolicity wins over
    the effect of singularities. We note that, in the planar case,
    hyperbolicity is exponential in nature, whereas the chopping up
    effect caused by singularities is only algebraic (in fact
    quadratic).} 
  (to be formulated below) the planar theory extends to the
  multidimensional case. That is to say, for the map, correlations of
  smooth functions decay 
  exponentially.
\end{statement}

Let us turn now to the complexity hypothesis, which is formulated for
the time-discretised process, the so-called Poincar\'e section map; see
Subsection~\ref{ssec:bill}. Billiards are singular dynamical systems, so if one
applies the billiard map $T$ to some nice, smooth convex (in
other words expanding) submanifold $\Sigma \subset \partial M$, then,
under the iterations of $T$, the images $T^n \Sigma$ of $\Sigma$ will
typically be chopped up into several smooth pieces (tangent collisions
or collisions at corner points of the boundary of the billiard table
indeed break the images of $\Sigma$). 

Let us denote by $K_n$ the upper bound on the number of smooth pieces
of $T^n \Sigma$.  The precise form of the so-called
subexponential complexity hypothesis is borrowed here from 
\cite[Subsection 2.4.2]{BT08}:

\begin{conjecture}[Complexity hypothesis]
For typical\footnote{In any reasonable sense of typicality, for
  instance, with respect to the $C^r$ topology ($r\ge 3$) on the
  scatterers.} three-dimensional billiards with smooth scatterers and 
finite horizon,  there exists $\lambda >1$, which is strict\-ly less
than the smallest expansion rate on the unstable cones of the
billiard, such that $K_n = o(\lambda^n)$. 
\end{conjecture}

We note here that B\'alint and T\'oth \cite{BT12} have also found an
example where the subexponential  complexity hypothesis does not
hold. This  example, however,  is quite special. Because of the utmost 
importance of settling the complexity hypothesis  in general, the
authors offer a bottle of Unicum Riserva\footnote{Unicum Riserva is a
  popular beverage specialty of the celebrated Hungarian firm Zwack
  Unicum PLC founded in 1790.} for the construction of a billiard which
exemplifies the complexity hypothesis\footnote{While it is known that
  finite-horizon periodic billiards with non-overlapping spheres exist
  in any dimension  \cite{H00, B02}, their explicit construction is far from
  straightforward, even in three dimensions. In particular, no such
  configuration can be obtained with lattice packings of spheres, that
  is, when congruent spherical scatterers are placed at the vertices
  of a regular lattice.
  The horizons
  of such billiards are infinite unless the scatterers are allowed to
  overlap \cite{S08}, which breaks the assumption of smooth
  scatterers.}.   

Let us make an additional remark in closing. Namely, there is an
interesting cultural difference between ergodic theory and other
branches of mathematics. For instance, in number theory, it is common
to prove conditional results, most notably assuming the validity of
Riemann  hypothesis. So far as we know, this is not common in the
theory of dynamical systems. To take the example of the theory of
multidimensional Sinai billiards (or perhaps of semi-dispersing
billiards as well), new developments might be achievable under the
complexity hypothesis. A convincing computational
evidence of its validity is also desirable.

\subsection{Some results for the random Lorentz and wind--tree
  processes}
\label{ssec:randwt}

Below we recall two recent closely connected results due to Lutsko and
T\'oth \cite{LT18,LT19}. They apply to two different processes in
$\bR^3$, the random Lorentz process, on the one hand, and the random
wind--tree process, on the other hand. The authors  prove that, in the
Boltzmann-Grad limit, both converge to Wiener processes. Our goal in
mentioning these results is to provide support to our conjectures
formulated in Subsection \ref{ssec:windconjectures}, since, here too,
one draws a close analogy between the Lorentz and the wind--tree
processes.  

\begin{remark}
  It is important to emphasize that this analogy applies to the random Lorentz
  and wind tree processes. For the periodic processes, details of the
  local dynamics are more relevant. In particular, the planar periodic
  wind--tree process has anomalous diffusive properties; see
  \cite{DHL14}. Another interesting model recently studied in
  \cite{ABB20} is a planar periodic wind--tree process with rounded off
  scatterers (or the wind tree process for a small extended hard
  discs). This process regains local hyperbolicity and it is expected
  to limit to Brownian motion, but potentially with a new type of
  scaling. 
\end{remark}

Let us introduce notations. Depending on the nature of the process,
the fixed scatterers are respectively \textit{balls} of radii $r$ or
upright \textit{cubes} of sides $r$. In both 
cases, the centers of the scatterers are fixed according to Poisson
point processes of positive density, $\rho > 0$. Initially, the point particle
is placed at the origin. In the Boltzmann-Grad limit, we have $r \to
0$ and $\rho \to \infty$ while (say) $\rho \, r^2 \to \pi$. The Lorentz process
$S_{r}^t: 0 \le t < \infty$ (resp. the wind--tree process $S_{r, [{\rm
p}]}^t: 0 \le t < \infty$) is a Markov jump process with a
probability tending to $1$. For more details, see \cite{LT18,LT19}.

The authors investigate in both papers the so-called
\textit{averaged-quen\-ched limit}, such that both the initial velocity
directions of the point particle and the configuration of the
scatterers are random and the limit is taken while averaging 
with respect to these random configurations.

In the forthcoming theorems, $T$ denotes a scale such
that\footnote{The difference in the conditions on the increase of $T$
  as $r \to 0$ in the two theorems is certainly a technical one.}, as
$r \to 0$, $T \to \infty$. We also let $\rho = \pi \, r^{-2}$.

\begin{theorem}[Random Lorentz process \cite{LT18}] 
  Assume, in addition, that $\lim_{r \to 0}r^2\,|\log r|^2 \,T = 0$. Then,
  as $r \to 0$, and in the aforementioned averaged-quenched sense, 
  \begin{equation*}
    \frac{S_r^{tT}}{\sqrt T} \Longrightarrow W(t)\qquad t \in [0,
    \infty)\,,
  \end{equation*} 
  where $W(t)$ is the standard Wiener process in $\bR^3$ and the
  convergence is weak convergence in $C[0, \infty)$.
\end{theorem}

\begin{theorem}[Random wind--tree process \cite{LT19}] 
  Assume, in addition, that $\lim_{r \to 0}r^2 \, T = 0$. Then, as $r
  \to 0$, and in the aforementioned averaged-quenched sense,
  \begin{equation*}
    \frac{S_{r,{p}}^{tT}}{\sqrt T} \Longrightarrow W(t)\qquad t
    \in [0, \infty)\,,
  \end{equation*}
  where $W(t)$ is the Wiener process  in $\bR^3$, with covariance
  matrix $\Sigma = {\rm diag} (v_1^2,v_2^2, v_3^2)$, defined in terms
  of the possible velocity directions, and the convergence is weak
  convergence in $C[0, \infty)$.
\end{theorem}

Before concluding this section, we mention a novel approach to random
Lorentz processes by  Aimino and Liverani \cite{AL18} via
deterministic walks in random environments.

\section{Fourier's Law of Heat Conduction}
\label{sec:fourier}

Ever since the formulation by Fourier of his \emph{th\'eorie analytique de la
  cha\-leur}, heat conduction has intrigued both engineers and
scientists, physicists and mathematicians alike, as has, in fact, much
of Fourier's scientific legacy \cite{Bernardin2019}. Besides the
articles \cite{Dhar2019Fourier,Bodineau2019,DeRoeck2019,Mendl2019}, 
which were published in the aforementioned volume,
many excellent additional surveys and volumes have been published,
recounting progress in this area; in particular and among the more
recent ones, see \cite{N99, BLR-B00, LLP03, D08, Lepri2016, Dhar2019}.

Among these reviews, Bonetto \emph{et al.} \cite{BLR-B00} has had
considerable influence in shaping current understanding of the topic
and focus research on key issues, depicting the state of affairs at
the turn of the millennium in a style accessible to both mathematics
and physics communities, covering both stochastic and deterministic
models. A more recent survey on the derivation of Fourier's law
starting from a Hamiltonian description, which is closely relevant to
our discussion, can be found in \cite{L18}. For the derivation of
large scale dynamics from stochastic models, a basic reference has
been and remains \cite{S91}.

While the pioneering work of B. J. Adler and T. E. Wainwright
\cite{Alder1957, Alder1959, Alder1960} established billiard models in
the form of hard spheres as potent tools for the numerical study of
nonequilibrium transport processes (as well as critical phenomena for
that matter), much of the theoretical effort on the study of transport
phenomena such as heat conduction has focused mostly on weakly
anharmonic chains which bear no connection to billiards; see,
e.g.,~\cite{Cowley1968}. This situation changed in the last two
decades with the emergence of new and promising models, which we turn
to below. In Subsection \ref{ssec:nelg}, we discuss the nonequilibrium
Lorentz gas and an interacting version of it, which, as we explain,
has features of billiard dynamics, but is not an actual billiard. The
next Subsection \ref{ssec:heatBLPS} deals with proper billiard models
of the type described in Subsection \ref{ssec:spatial}. In a regime of
rare interactions, these billiards lend themselves to a two-step
strategy to deriving Fourier's law of heat conduction. The main ideas
are recalled and perspectives given.

\subsection{Nonequilibrium Lorentz gases}
\label{ssec:nelg}

Before we go into our main discussion, we wish to mention
an interesting model of heat conduction thought to obey Fourier's law,
namely a form of periodic Lorentz gas whose circular scatterers freely
rotate and exchange energy with point particles via ``perfectly rough
collisions'', thereby mediating interactions among the gas particles
\cite{M-MCL01, LLM-M03}. While this model is not quite a billiard, it
has some of its main features. 

To motivate the model, let us mention that a two-dimensional slab
(i.e.~an array of cells with finite horizontal and vertical lengths) of
the periodic Lorentz gas may be driven out of equilibrium by putting, say,
its vertical boundaries in contact with external reservoirs or
baths. These reservoirs could be chemostats, acting like absorbing
boundaries for particles that collide with them, and randomly
injecting particles into the system at specified rates. Or they may be
thermal baths which reflect colliding particles while randomly
changing their velocities according to set temperatures. And a 
combination of both chemostats and thermal baths is also possible.

To give a concrete example, we mention that a numerical study of such
a Lorentz gas driven out of equilibrium by 
two thermal baths at different temperatures was reported in
\cite{Alonso1999}, where, under reasonable assumptions, it was found
that temperature profiles are consistent with macroscopic laws. In
such a situation, the distributions of the gas particles exhibit near
local equilibrium properties. That is, the velocity distributions are
``close'' to Maxwellians with a  
position-dependent temperature (how close depends on the value of
the local temperature gradient). This is sometimes referred to as \emph{local
thermal equilibrium}. Of course, in this system, particles do not
interact, so that each one of them retains the same energy between
successive interactions with the thermal baths. Local thermal
equilibrium is therefore merely a reflection of the superposition of
independent gas particles. Loosely speaking, ``warm'' particles stay
warm, just as ``cold'' particles stay cold; the answer to the question
of where and in what proportions they mix determines the local
temperature. One may be critical of the lack of actual equilibration
among the particles sharing the same location (meaning lattice cell),
but that is the essence of the model and one cannot have it both ways.   

The Lorentz gas with rotating scatterers referred to earlier is
precisely designed to achieve an exchange of energy between particles
sharing the same location that the usual Lorentz gas lacks. Thanks to
the mechanism of perfectly rough collisions proposed by the authors of
\cite{LLM-M03} (see, however, \cite{Rateitschak2000} for a similar
proposition), a particle which enters a cell will effectively leave
the cell with an energy different from that with which it entered,
which, in a sense, behaves like a random variable similar to that
generated by a thermal bath at the local temperature of the cell
(which is therefore a dynamically evolving quantity itself). It must
be acknowledged that our description of the phenomenology of this
model is a bit of an idealisation. Reality is indeed more intricate, but
such details are not relevant here. 

Thus why is this model not a billiard? This has to do with the
mechanism of perfectly rough collisions alluded to above. 
Circular scatterers, whose centers are affixed to the vertices of the
lattice possess a rotational degree of freedom. Whenever a
point particle of the gas hits a scatterer, the collision rules
are such that the radial component of the particle's momentum changes
sign (in the same way that it does in an elastic collision). At the same time,
the tangential component of its momentum and the angular momentum of
the rotating scatterer undergo a linear interaction parametrised by
the equivalent of a friction coefficient, but such that the total
energy is fixed. This is therefore different from an elastic collision,
which, irrespective of the rotational degree of freedom of the
scatterers, would leave the energy of the particle
unchanged\footnote{It is worthwhile noting here that a different model
with similar properties was considered in \cite{Eckmann2006}, where
circular scatterers were replaced by rods (termed ``needles'' by the
authors because they have zero thickness). The energy exchange
mechanism between the rotating rods and point particles is defined  
through conservation of linear and angular momenta. Contrary to the
perfectly rough collisions described above, such collisions are in
fact consistent with billiard dynamics \cite{GGM-MSzXX}.}. 

Regardless of the fact that its dynamics is not Hamiltonian, the model
\cite{LLM-M03} gave rise to a number of interesting contributions; see
in particular \cite{EY06}.

\subsection{Heat conduction in gases of locally confined hard balls}
\label{ssec:heatBLPS}

Even if the models described above were Hamiltonian, as is the variant
with rotating rods numerically studied in \cite{Eckmann2006}, a
systematic derivation of Fourier's law remains far out of reach. The
control of correlations is indeed very  difficult to achieve due to the
high-dimensionality of the models.  

In that respect, the billiard models studied \cite{Gaspard2008}
belong to the class of BLPS models \cite{BLPS92} described in Section
\ref{ssec:BLPS}. While these models are high-dimensional billiards
with many of the properties of hard ball gases, the trapping mechanism
of the individual balls induces a spatial order which considerably
simplifies the physical picture in the sense that mass transport is
prohibited and only energy transport takes place. This has opened new 
perspectives for a systematic derivation of Fourier's law, especially
in a regime of rare interactions.

\subsubsection{Rarely interacting BLPS billiards}
\label{ssec:rarely}

To explain the notion of rare interactions, we note that BLPS models
have two readily identifiable timescales. The first one has to do with
the trapping mechanism and denotes the average time that separates
successive collisions between a moving ball and the walls of its cell
(assuming no interactions with a neighbouring moving ball takes place
in that time interval). Call $\tw$ this time, which
trivially scales with the inverse square root of the kinetic energy of
the ball and otherwise depends only on the geometry of the cell
through the sum $\rhom + \rhof$ of the radii of fixed scatterers
$\rhof$ and moving balls $\rhom$; see \cite{Chernov1997}. The second 
timescale is that which, in average, separates successive collisions
between two neighbouring moving balls (under the assumption that no
interactions involving other balls take place in that time
interval). Call $\tb$ this time, which trivially scales with the 
inverse square root of the sum of the kinetic energies of the two
balls. Importantly, its dependence on the geometry of the cells
involves a parameter, that which measures how close the radius $\rhom$
of the moving balls is to the critical radius $\rhoc$ defined in Subsection
\ref{ssec:BGNST}. 

Crucially, one can vary $\rhom$ and $\rhof$ while keeping their sum,
and hence $\tw$, fixed, even as $\rhom \to \rhoc$. Assuming $\epsilon
:= \rhom - \rhoc \,:\, 0 <  \epsilon \ll 
1$, it is not difficult to show that $\tb \propto \epsilon^{-1}$
\cite{GG08}. We therefore have the following separation between the
two timescales, 
\begin{equation*}
  \tb \gg \tw\,.
\end{equation*}
This result embodies the rare interaction regime and implies a
typical relaxation to local equilibrium of the internal degrees of
freedom (averaging of the positions and velocity directions) 
\emph{before} a collision among moving balls takes place. It is key
to reducing the dynamics of energy exchanges among moving balls to a
Markov jump process; see below. 

This discussion naturally brings us back to an important aspect
of the rarely interacting BLPS billiards, which concerns the
distinction between microscopic and macroscopic observables alluded to
in Subsection \ref{ssec:increase}. Much like with diffusion in
the Machta-Zwanzig regime of finite-horizon periodic Lorentz gases,
the lattice coordinates provide the natural basis for inferring a set of
macroscopic positions with attached local energies given in terms of
the velocities squared of the trapped particles. These local
energies can furthermore be interpreted as temperatures, which
are indeed defined without ambiguity under the assumption of
relaxation to local equilibria, expected in the regime of rare
interactions. Their counterparts, i.e.~the microscopic variables, which
represent the internal degrees of freedom (positions of the balls
inside the cells and their velocity angles) are rapidly averaged over
on the timescales of energy exchanges and do not bear further
influence on the process. 

We remark that, as is the case with the Machta-Zwanzig regime of
diffusion discussed in Subsection \ref{ssec:random}, the
dimensionality of the dynamics has little relevance to the assumptions
that drive the local relaxation process described above. Billiard
models of both higher and lower dimensionalities have in fact been
studied elsewhere \cite{Gaspard2012,  BGNSzT17}, exhibiting similar
properties.

\subsubsection{Reduction to a Markov jump process}
\label{ssec:EnergyJump}

A minimal model to study the process of energy exchanges in the rare
interaction regime is a billiard table of only two cells, each with
its own moving ball trapped inside it, and sharing a boundary through
which collisions between the two balls are possible. At the value $\rhom =
\rhoc$, i.e.~$\epsilon = 0$ in the notation introduced above, the two
balls are effectively isolated and cannot collide; the model is
then akin to an insulator. This four-dimensional billiard is
therefore the product of two Sinai-billiards whose cells have
triangular shapes such as outlined by the dotted circles in Figure
\ref{fig:blps}.

One would like to study the $0 < \epsilon \ll 1$ energy-conducting
regime of rare interactions as a perturbation of this insulating
regime\footnote{It is indeed an interesting feature of these models
  that a normally-conducting regime is expected to emerge out of an
  insulating one. The situation is very different with weakly
  anharmonic oscillators, whose harmonic limits have infinite
  conductivity.}. Theoretically, this would amount to generalising the
method of standard pairs \cite{ChD07} to this situation. A somewhat
similar strategy was actually described in \cite{PajorGyulai2010} to
study the diffusive limit of a gas of two hard 
balls moving in a finite-horizon dispersing planar billiard table. 

As explained in \cite{PajorGyulai2012}, however,
\begin{quote}
  At the present state of dynamical methods [this problem]
  unfortunately defies a rigorous approach. So far the 
  apparently strongest method: that of standard pairs elaborated in
  detail in \cite{ChD07} does not permit extension to truly higher-dimensional
  models, like this one. (The main technical reason is that the
  conservation of the standard pair structure after one collision can
  not be controlled for some particular configurations of the
  colliding disks.) 
\end{quote}
With respect to the model under consideration, it implies, for
instance, that there is no way of bounding the correlation decay of
nice functions that would be necessary to justify the first step of
the Gaspard-Gilbert two-step approach outlined in \cite{Gaspard2008}. 

It was with this obstacle in mind that a simpler variant of the BLPS
class of models was designed, as described in  Subsection \ref{ssec:BGNST},
which, in its minimal configuration is a three-dimensional
semi-dispersing billiard. In \cite{BGNSzT17}, the reduction to a
Markov jump process was discussed 
on phenomenological grounds and a derivation of the associated
transition kernel was given in terms of conditional mean free times.
Still, progress towards applying the method of \cite{ChD07} to this
model so far remains too limited to make any serious claim.
Nevertheless, the authors of \cite{Balint2018} have announced a result
on this issue and have, in fact, already clarified one important step
toward a rigorous proof.

Along these lines, we formulate below a conjecture which experts will
have anticipated, but which gives us the opportunity to emphasize
a result of fundamental importance\footnote{We should, in particular,
  note that the content of this conjecture is a necessary step towards
  controlling the decay of correlations between neighbouring particles
  in the model of heat conduction presented in
  section~\ref{ssec:BGNST}. The explanation is simple: the minimal
  three-dimensional billiard model isomorphic to a single ball-piston
  pair has corner points, a case not covered by
  reference~\cite{BDL18} (or by reference~\cite{Balint2018} for that
  matter).}:  

\begin{conjecture}
  The main result of \cite{BDL18} on the exponential decay of
  correlations of planar Sinai billiard flows is also true for billiards
  with corner points (assuming that the angle of the intersecting
  boundary pieces at these corners is not zero).   
\end{conjecture}

To be more specific, we recall that, as said in Subsection
\ref{ssec:Markov}, a strong advantage of the method of standard pairs  
is that it is also suitable for the joint treatment of perturbations
of a dynamical system, a situation which, by its very nature, arises
in the rare interaction regime of the models under consideration. 
Building upon the breakthrough results of \cite{BDL18}, the
authors of \cite{Balint2018} thus showed that, if one considers a
dispersing billiard flow on the  two-dimensional torus whose initial
measure is concentrated on the highly singular measure determined by a
standard pair, then one still has exponential convergence to the
absolutely continuous equilibrium Liouville measure. 

\subsubsection{Hydrodynamic limit}
\label{ssec:hydro}

Even if the strategy outlined above for the reduction of the energy
exchange dynamics to a Markov jump process were fully justified, the
problem of deriving Fourier's law of heat conduction (and determining
the heat conductivity) is a separate challenge and, at that, rather more
complicated than obtaining the diffusive limits of simple symmetric 
random walks such as discussed in Subsection \ref{ssec:random}.

In order to implement the second step of the Gaspard-Gilbert 
approach \cite{Gaspard2008}  for calculating the heat conductivity of the
BGNST model of \cite{BGNSzT17} (as well as that of other
models\footnote{A computational method which relies on Varadhan's
  variational formula and the calculation of exact bounds was
  implemented in \cite{GG17} and applied to a 
  specific model, yielding a precise estimate of heat conductivity
  with a (negative) contribution from dynamical correlations estimated
  to be less than four parts in ten thousands of the static
  contribution---the published figure is in fact much more precise.}),
one ought to take  the hydrodynamic limit of the associated Markov
jump process. At 
present the most hopeful approach is to apply the  variational method
of Varadhan \cite{Varadhan1993}; see also \cite{S91} for a general
discussion in the framework of stochastic lattice gases.

An important contribution in that direction is a result by Sasada
\cite{Sasada2015}, who obtained a lower bound on the spectral gap of
the Gaspard-Gilbert model \cite{Gaspard2008}, namely
$\mathrm{const.}\,N^{-2}$ (with a strictly positive constant),  where 
$N$ denotes the size of the system. As far as the BGNST model is
concerned, the problem is slightly more complicated due to the hybrid
nature of ball-piston cells. We transpose Sasada's result to the following

\begin{conjecture}[Spectral gap for the BGNST model of \cite{BGNSzT17}]
  The spectral gap of the piston model is bounded below by
  $\mathrm{const.}\,N^{-2}$ (with a strictly positive constant),  where
  $N \in \mathbb Z^+$ denotes the number of ball-piston pairs (aligned
  along a one-dimensional chain).  
\end{conjecture}

Although Varadhan's method was formulated for models different from
ours, it is expected to work here as well. There is, however, a
caveat, namely that Varadhan's conditions on the coefficients of his 
models are perhaps too restrictive for the BGNST model. Consequently,
more work is needed so as to relax these conditions and make them
applicable.

\section*{Acknowledgement} 

DSz wishes to use this occasion to express his utmost gratitude to his
late friend Andr\'as Kr\'amli (1943--2019). As a graduate student,
Andr\'as worked under the supervision Ya G Sinai and thus learnt the
theory of smooth dynamical systems. His knowledge and many years of
collaboration were essential in helping DSz to understand Sinai's
work on billiards and gain deeper understanding of this wonderful
theory. This is also an appropriate point for DSz to express his
gratefulness for Yakov  Grigoryevich Sinai's continuous attention and
support. 

The authors also express their gratitude to Vered Rom-Kedar, Sylvain
Crovisier, Vadim Kaloshin and Marco Lenci for their most helpful feedback.

\appendix
\section{Some results on wind--tree models}
\label{sec:Ahom}

Below we recall two theorems which we refer to in Subsection
\ref{ssec:infty}. Both models were introduced in Subsection
\ref{ssec:wind}. The first theorem claims (infinite measure)
ergodicity about an aperiodic model with square scatterers. 

\begin{theorem}
  \label{thm:M-ST}
  [M\'alaga-Sabogal--Troubetzkoy \cite{M-ST16}] 
  Using the notations of Section \ref{sec:aperwindtree},
  we
  claim: There is a dense subset $\mG \subset \mA^{\bZ^{2}}$ of
  parameters such that for each $g \in \mG$ there 
  is a dense $G_\delta-$subset of directions $\mH \subset \bS_{1}$
  of full measure such that the billiard flow on $Q_g$ in the direction
  $\theta$ is ergodic for every $\theta \in \mH$.
\end{theorem}

For simplicity we recall only part of a theorem of \cite{DHL14} that
reflects well its flavour. It claims that the wind--tree process with
periodic rectangular scatterers of size $a \times b$ makes large
excursions (cf. {escape of local information to infinity} in
reference to \cite{GLA75}).

\begin{theorem}\label{thm:DHL}[Delecroix-Hubert-Leli\`evre
  \cite{DHL14}] For Lebesgue-almost $(a, b) \in [0, 1)^2$, for
  Lebesgue-almost all $\theta$ and every point in $Q = Q_{a, b}$ (with
  infinite forward orbit)
  \begin{equation*}
    \limsup_{T \to \infty} \frac{\log d(p, S^{T}_{\theta}(p))}{\log T} =
    \frac{2}{3}\,.
  \end{equation*}
\end{theorem}


\begin{thebibliography}{GGMMS20}

\bibitem[AA67]{Arnold1967}
Vladimir~I Arnold and Andr{\'e} Avez.
\newblock {\em Probl\`emes Ergodiques de la M\'ecanique Classique}.
\newblock Gauthier--Villars, Paris, 1967.

\bibitem[AACG99]{Alonso1999}
Daniel Alonso, Roberto Artuso, Giulio Casati, and Italo Guarneri.
\newblock Heat conductivity and dynamical instability.
\newblock {\em Physical Review Letters}, 82:1859, March 1999.
\newblock \url{https://link.aps.org/doi/10.1103/PhysRevLett.82.1859}.

\bibitem[ABB20]{ABB20}
Hassan Attarchi, Mark Bolding, and Leonid~A Bunimovich.
\newblock Ehrenfests’ wind--tree model is dynamically richer than the lorentz
  gas.
\newblock {\em Journal of Statistical Physics}, 180:440--458, 2020.
\newblock \url{https://link.springer.com/article/10.1007/s10955-019-02460-8}.

\bibitem[AL20]{AL18}
Romain Aimino and Carlangelo Liverani.
\newblock Deterministic walks in random environment.
\newblock {\em The Annals of Probability}, 48(5):2212--2257, 09 2020.
\newblock \url{https://projecteuclid.org/euclid.aop/1600826470}.

\bibitem[Ano67]{A67}
Dmitry~Victorovich Anosov.
\newblock Geodesic flows on closed riemann manifolds with negative curvature.
\newblock {\em Proc Steklov Inst Math}, 90(5):3--210, 1967.
\newblock \url{http://mi.mathnet.ru/eng/tm2795}.

\bibitem[AS67]{AS67}
Dmitry~V Anosov and Yakov~G Sinai.
\newblock Some smooth ergodic systems.
\newblock {\em Russian Mathematical Surveys}, 22(5):103--167, 1967.
\newblock
  \url{https://iopscience.iop.org/article/10.1070/RM1967v022n05ABEH001228}.

\bibitem[Ash15]{Ashley2015}
Steven Ashley.
\newblock Core concept: Ergodic theory plays a key role in multiple fields.
\newblock {\em Proceedings of the National Academy of Sciences},
  112(7):1914--1914, 2015.
\newblock \url{https://www.pnas.org/content/112/7/1914}.

\bibitem[AW57]{Alder1957}
Berni~J Alder and Thomas~E Wainwright.
\newblock Phase transition for a hard sphere system.
\newblock {\em The Journal of Chemical Physics}, 27(5):1208--1209, 1957.
\newblock \url{https://aip.scitation.org/doi/10.1063/1.1743957}.

\bibitem[AW59]{Alder1959}
Berni~J Alder and Thomas~E Wainwright.
\newblock Studies in molecular dynamics. {I}. general method.
\newblock {\em The Journal of Chemical Physics}, 31(2):459--466, 1959.
\newblock \url{https://aip.scitation.org/doi/10.1063/1.1730376}.

\bibitem[AW60]{Alder1960}
Berni~J Alder and Thomas~E Wainwright.
\newblock Studies in molecular dynamics. {II}. behavior of a small number of
  elastic spheres.
\newblock {\em The Journal of Chemical Physics}, 33(5):1439--1451, 1960.
\newblock \url{https://aip.scitation.org/doi/10.1063/1.1731425}.

\bibitem[Bal00]{B00}
Viviane Baladi.
\newblock {\em Positive Transfer Operators and Decay of Correlations}.
\newblock Wolrd Scientific, Singapore, 2000.
\newblock \url{https://www.worldscientific.com/doi/abs/10.1142/3657}. Erratum
  available at \url{http://baladi.perso.math.cnrs.fr/erratum.pdf}.

\bibitem[BCST02]{BNSzT02}
P{\'e}ter B{\'a}lint, Nikolai Chernov, Domokos Sz{\'a}sz, and Imre~P{\'e}ter
  T{\'o}th.
\newblock Multi--dimensional semi--dispersing billiards: singularities and the
  fundamental theorem.
\newblock {\em Annales Henri Poincar\'e}, 3(3):451--482, 2002.
\newblock \url{https://link.springer.com/article/10.1007/s00023-002-8624-7}.

\bibitem[BCST03]{BChSzT03}
P{\'e}ter B{\'a}lint, Nikolai Chernov, Domokos Sz{\'a}sz, and Imre~P{\'e}ter
  T{\'o}th.
\newblock Geometry of multi--dimensional dispersing billiards.
\newblock {\em Ast\'erisque}, 286:119--150, 2003.
\newblock \url{http://www.numdam.org/item/AST_2003__286__119_0}.

\bibitem[BDL18]{BDL18}
Viviane Baladi, {Mark F} Demers, and Carlangelo Liverani.
\newblock Exponential decay of correlations for finite horizon {S}inai billiard
  flows.
\newblock {\em Inventiones mathematicae}, 211(1):39--177, 2018.
\newblock \url{https://link.springer.com/article/10.1007/s00222-017-0745-1}.

\bibitem[BF19]{Bernardin2019}
C\'edric Bernardin and Patrick Flandrin.
\newblock Foreword: {F}ourier and the science of today.
\newblock {\em Comptes Rendus Physique}, 20(5):387--391, 2019.
\newblock
  \url{http://www.sciencedirect.com/science/article/pii/S1631070519301161}.

\bibitem[BFK06]{Berkovitz2006}
Joseph Berkovitz, Roman Frigg, and Fred Kronz.
\newblock The ergodic hierarchy, randomness and {H}amiltonian chaos.
\newblock {\em Studies in History and Philosophy of Science Part B: Studies in
  History and Philosophy of Modern Physics}, 37(4):661--691, 2006.
\newblock
  \url{http://www.sciencedirect.com/science/article/pii/S1355219806000700}.

\bibitem[BGN{\etalchar{+}}17]{BGNSzT17}
P{\'e}ter B{\'a}lint, Thomas Gilbert, P{\'e}ter N{\'a}ndori, Domokos Sz{\'a}sz,
  and Imre~P{\'e}ter T{\'o}th.
\newblock On the limiting {M}arkov process of energy exchanges in a rarely
  interacting ball--piston gas.
\newblock {\em Journal of Statistical Physics}, 166(3):903--925, 2017.
\newblock \url{https://link.springer.com/article/10.1007/s10955-016-1598-5}.

\bibitem[BGSR19]{Bodineau2019}
Thierry Bodineau, Isabelle Gallagher, and Laure Saint-Raymond.
\newblock A microscopic view of the fourier law.
\newblock {\em Comptes Rendus Physique}, 20(5):402--418, 2019.
\newblock
  \url{http://www.sciencedirect.com/science/article/pii/S1631070519301082}.

\bibitem[Bir31]{B31}
George~D Birkhoff.
\newblock Proof of the ergodic theorem.
\newblock {\em Proceedings of the National Academy of Sciences},
  17(12):656--660, 1931.
\newblock \url{https://www.pnas.org/content/17/12/656}.

\bibitem[BK32]{BK32}
George~D Birkhoff and Bernard~O Koopman.
\newblock Recent contributions to the ergodic theory.
\newblock {\em Proceedings of the National Academy of Sciences},
  18(3):279--282, 1932.
\newblock \url{https://www.pnas.org/content/18/3/279}.

\bibitem[BLPS92]{BLPS92}
Leonid~A Bunimovich, Carlangelo Liverani, Alessandro Pellegrinotti, and Yurii~M
  Suhov.
\newblock Ergodic systems of $n$ balls in a billiard table.
\newblock {\em Communications in Mathematical Physics}, 146:357--396, May 1992.
\newblock \url{https://link.springer.com/article/10.1007/BF02102633}.

\bibitem[BLRB00]{BLR-B00}
Federico Bonetto, Joel~L Lebowitz, and Luc Rey-Bellet.
\newblock Fourier's law: a challenge to theorists.
\newblock In Athanasios~S Fokas, Alexander Grigoryan, Thomas W~B Kibble, and
  Boguslaw Zegarlinski, editors, {\em Mathematical Physics 2000}, chapter~8,
  pages 128--150. Imperial College Press, London, 2000.
\newblock
  \url{http://www.worldscientific.com/doi/abs/10.1142/9781848160224_0008}.

\bibitem[BNST18]{Balint2018}
Péter Bálint, Péter Nándori, Domokos Szász, and Imre~Péter Tóth.
\newblock Equidistribution for standard pairs in planar dispersing billiard
  flows.
\newblock {\em Annales Henri Poincaré}, 19(4):979--1042, April 2018.
\newblock \url{https://link.springer.com/article/10.1007/s00023-018-0648-8}.

\bibitem[BPP{\etalchar{+}}85]{BPPSS85}
Carlo Boldrighini, Alessandro Pellegrinotti, Errico Presutti, Yakov~G Sinai,
  and Mikhail~R Soloveichik.
\newblock Ergodic properties of a semi--infinite one--dimensional system of
  statistical mechanics.
\newblock {\em Communications in Mathematical Physics}, 101(3):363--382,
  September 1985.
\newblock \url{https://link.springer.com/article/10.1007/BF01216095}.

\bibitem[BS73]{BS73}
Leonid~A Bunimovich and Yakov~G Sinai.
\newblock On a fundamental theorem in the theory of dispersing billiards.
\newblock {\em Mathematics of the {USSR}--Sbornik}, 19(3):407--423, apr 1973.
\newblock
  \url{https://iopscience.iop.org/article/10.1070/SM1973v019n03ABEH001786}.

\bibitem[BS80]{BS80}
Leonid~A Bunimovich and Yakov~G Sinai.
\newblock Markov partitions for dispersed billiards.
\newblock {\em Communications in Mathematical Physics}, 78:247--280, 1980.
\newblock \url{https://link.springer.com/article/10.1007/BF01942372}.

\bibitem[BS81]{BS81}
Leonid~A Bunimovich and Yakov~G Sinai.
\newblock Statistical properties of {L}orentz gas with periodic configuration
  of scatterers.
\newblock {\em Communications in Mathematical Physics}, 78(4):479--497, 1981.
\newblock \url{https://link.springer.com/article/10.1007/BF02046760}.

\bibitem[BS86]{BS86}
Leonid~A Bunimovich and Yakov~G Sinai.
\newblock Markov partition for dispersed billiards.
\newblock {\em Communications in Mathematical Physics}, 107(2):357--358, June
  1986.
\newblock \url{https://link.springer.com/article/10.1007/BF01209400}.

\bibitem[BS96]{BS96}
Leonid~A Bunimovich and Herbert Spohn.
\newblock Viscosity for a periodic two disc fluid: an existence proof.
\newblock {\em Communications in Mathematical Physics}, 176(3):661--680, 1996.
\newblock \url{https://link.springer.com/article/10.1007/BF02099254}.

\bibitem[BSC90]{BChS90}
Leonid~A Bunimovich, Yakov~G Sinai, and Nikolai Chernov.
\newblock Markov partitions for two--dimensional hyperbolic billiards.
\newblock {\em Russian Mathematical Surveys}, 45(3):105--152, 1990.
\newblock
  \url{https://iopscience.iop.org/article/10.1070/RM1990v045n03ABEH002355}.

\bibitem[BSC91]{BChS91}
Leonid~A Bunimovich, Yakov~G Sinai, and Nikolai Chernov.
\newblock Statistical properties of two--dimensional hyperbolic billiards.
\newblock {\em Russian Mathematical Surveys}, 46(4):47--106, 1991.
\newblock
  \url{https://iopscience.iop.org/article/10.1070/RM1991v046n04ABEH002827}.

\bibitem[BT02]{B02}
K\'aroly {B\"{o}r\"{o}czky Jr} and G\'abor Tardos.
\newblock The longest segment in the complement of a packing.
\newblock {\em Mathematika}, 49(1-2):45--49, 2002.
\newblock
  \url{https://londmathsoc.onlinelibrary.wiley.com/doi/abs/10.1112/S002557930001603X}.

\bibitem[BT08]{BT08}
P{\'e}ter B{\'a}lint and Imre~P{\'e}ter T{\'o}th.
\newblock Exponential decay of correlations in multi--dimensional dispersing
  billiards.
\newblock {\em Annales Henri Poincar\'e}, 9(7):1309--1369, 2008.
\newblock \url{https://link.springer.com/article/10.1007/s00023-008-0389-1}.

\bibitem[BT12]{BT12}
P{\'e}ter B{\'a}lint and Imre~P{\'e}ter T{\'o}th.
\newblock Example for exponential growth of complexity in a finite horizon
  multi--dimensional dispersing billiard.
\newblock {\em Nonlinearity}, 25(5):1275--1297, May 2012.
\newblock \url{https://iopscience.iop.org/article/10.1088/0951-7715/25/5/1275}.

\bibitem[Bun19]{B19}
Leonid~A Bunimovich.
\newblock Billiards.
\newblock In Helge Holden and Ragni Piene, editors, {\em The Abel Prize
  2013--2017}, pages 195--205. Springer, Cham, 2019.
\newblock \url{https://link.springer.com/chapter/10.1007/978-3-319-99028-6_7}.

\bibitem[CD09]{ChD07}
Nikolai Chernov and Dmitry Dolgopyat.
\newblock {\em Brownian Brownian Motion--I}, volume 198.
\newblock Memoirs of the American Mathematical Society, Providence, RI, 2009.
\newblock \url{https://www.ams.org//books/memo/0927/}.

\bibitem[CFS82]{CFS82}
Isaak~P Cornfeld, Sergei~V Fomin, and Yakov~G Sinai.
\newblock {\em Ergodic Theory}.
\newblock Springer Verlag, New York, 1982.
\newblock \url{http://link.springer.com/book/10.1007/978-1-4615-6927-5}.

\bibitem[Che82]{Chernov1982}
Nikolai Chernov.
\newblock Structure of transversal leaves in multidimensional semidispersing
  billiards.
\newblock {\em Functional Analysis and Its Applications}, 16(4):270--280,
  October 1982.
\newblock \url={https://link.springer.com/article/10.1007/BF01077849}.

\bibitem[Che97]{Chernov1997}
Nikolai Chernov.
\newblock Entropy, {L}yapunov exponents, and mean free path for billiards.
\newblock {\em Journal of Statistical Physics}, 88(1--2):1--29, 1997.
\newblock \url{https://link.springer.com/article/10.1007/BF02508462}.

\bibitem[CM06]{ChM06}
Nikolai Chernov and Roberto Markarian.
\newblock {\em Chaotic Billiards}, volume 127 of {\em Mathematical Surveys and
  Monographs}.
\newblock American Mathematical Society, Providence, RI, 2006.
\newblock \url{https://bookstore.ams.org/surv-127}.

\bibitem[Cow68]{Cowley1968}
Roger~A Cowley.
\newblock Anharmonic crystals.
\newblock {\em Reports on Progress in Physics}, 31(1):123--166, jan 1968.
\newblock \url{https://iopscience.iop.org/article/10.1088/0034-4885/31/1/303}.

\bibitem[Det12]{D12}
Carl~P Dettmann.
\newblock New horizons in multidimensional diffusion: The {L}orentz gas and the
  riemann hypothesis.
\newblock {\em Journal of Statistical Physics}, 146:181--204, November 2012.
\newblock \url{https://link.springer.com/article/10.1007/s10955-011-0397-2}.

\bibitem[DH19]{DeRoeck2019}
Wojciech {De Roeck} and Fran{\c c}ois Huveneers.
\newblock Glassy dynamics in strongly anharmonic chains of oscillators.
\newblock {\em Comptes Rendus Physique}, 20(5):419--428, 2019.
\newblock
  \url{http://www.sciencedirect.com/science/article/pii/S1631070519301136}.

\bibitem[Dha08]{D08}
Abhishek Dhar.
\newblock Heat transport in low--dimensional systems.
\newblock {\em Advances in Physics}, 57(5):457--537, September 2008.
\newblock \url{https://www.tandfonline.com/doi/abs/10.1080/00018730802538522}.

\bibitem[DHL14]{DHL14}
Vincent Delecroix, Pascal Hubert, and Samuel Lelievre.
\newblock Diffusion for the periodic wind--tree model.
\newblock {\em Annales Scientifiques de L'École Normale Superieure},
  47(6):1085--1110, 2014.
\newblock
  \url{https://smf.emath.fr/publications/diffusion-du-vent-dans-les-arbres}.

\bibitem[DKK19]{Dhar2019}
Abhishek Dhar, Anupam Kundu, and Aritra Kundu.
\newblock Anomalous heat transport in one dimensional systems: A description
  using non-local fractional-type diffusion equation.
\newblock {\em Frontiers in Physics}, 7:159, 2019.
\newblock \url{https://www.frontiersin.org/article/10.3389/fphy.2019.00159}.

\bibitem[DL91]{DL91}
Victor Donnay and Carlangelo Liverani.
\newblock Potentials on the two--torus for which the {H}amiltonian flow is
  ergodic.
\newblock {\em Communications in Mathematical Physics}, 135(2):267--302,
  January 1991.
\newblock \url{https://link.springer.com/article/10.1007/BF02098044}.

\bibitem[Dob94]{D94}
Roland~L Dobrushin.
\newblock A mathematical approach to foundations of statistical mechanics.
\newblock Preprint ESI 179, 12 1994.
\newblock Scientific Publications of Erwin Schr\"odinger Institute (ESI),
  \url{https://www.esi.ac.at/static/esiprpr/esi179.pdf}.

\bibitem[Don51]{D51}
Monroe~David Donsker.
\newblock An invariance principle for certain probability limit theorems.
\newblock In {\em Four Papers on Probability}, 6, pages 50--58. Memoirs of the
  American Mathematical Society, Providence, RI, 1951.
\newblock \url{https://bookstore.ams.org/memo-1-6/}.

\bibitem[Dru00]{D00}
Paul Drude.
\newblock Zur elektronentheorie der metalle.
\newblock {\em Annalen der Physik}, 306(3):566--613, 1900.
\newblock
  \url{https://onlinelibrary.wiley.com/doi/abs/10.1002/andp.19003060312}.

\bibitem[DS19]{Dhar2019Fourier}
Abhishek Dhar and Herbert Spohn.
\newblock Fourier's law based on microscopic dynamics.
\newblock {\em Comptes Rendus Physique}, 20(5):393--401, 2019.
\newblock
  \url{http://www.sciencedirect.com/science/article/pii/S1631070519301100}.

\bibitem[DSV08]{Dolgopyat2008}
Dmitrii Dolgopyat, Domokos Sz{\'a}sz, and Tam{\'a}s Varj{\'u}.
\newblock Recurrence properties of planar {L}orentz process.
\newblock {\em Duke Mathematical Journal}, 142(2):241--281, 04 2008.
\newblock \url{https://projecteuclid.org/euclid.dmj/1206642155}.

\bibitem[DSV09]{Dolgopyat2009}
Dmitrii Dolgopyat, Domokos Sz{\'a}sz, and Tam{\'a}s Varj{\'u}.
\newblock Limit theorems for locally perturbed planar {L}orentz processes.
\newblock {\em Duke Mathematical Journal}, 148(3):459--499, May 2009.
\newblock \url{https://projecteuclid.org/euclid.dmj/1245350754}.

\bibitem[Dui20]{Democritus}
Brian Duignan.
\newblock Democritus.
\newblock Encyclop{\oe}dia Britannica,
  \url{https://www.britannica.com/biography/Democritus}, 2020.
\newblock Accessed: 2020/05/31.

\bibitem[Dup05]{D06}
Bertrand Duplantier.
\newblock Brownian motion, “diverse and undulating”.
\newblock In Thibault Damour, Olivier Darrigol, Bertrand Duplantier, and
  Vincent Rivasseau, editors, {\em Einstein, 1905--2005}, chapter~8, pages
  201--293. Birkh\"auser, Basel, 2005.
\newblock \url{https://link.springer.com/chapter/10.1007/3-7643-7436-5_8}.

\bibitem[EEA14]{EE12}
Paul Ehrenfest and Tatiana Ehrenfest-Afanassjewa.
\newblock Begriffliche grundlagen der statistischen auffassung in der mechanik.
\newblock In Felix Klein and Conrad~Heinrich M\"uller, editors, {\em
  Encyklop\"adie der mathematischen wissenschaften, IV Mechanik}, 32, pages
  773--860. Vieweg+Teubner Verlag, Wiesbaden, 1907--1914.
\newblock \url{https://link.springer.com/chapter/10.1007/978-3-663-16028-1_11}.

\bibitem[Ein05]{E05}
Albert Einstein.
\newblock \"uber die von der molekularkinetischen theorie der w\"arme
  geforderte bewegung von in ruhenden fl\"ussigkeiten suspendierten teilchen.
\newblock {\em Annalen der Physik}, 322(8):549--560, 1905.
\newblock
  \url{https://onlinelibrary.wiley.com/doi/abs/10.1002/andp.19053220806}.

\bibitem[EK46]{EK46}
Paul Erd{\"o}s and Mark Kac.
\newblock On certain limit theorems of the theory of probability.
\newblock {\em Bulletin of the American Mathematical Society}, 52(4):292--303,
  October 1946.
\newblock
  \url{https://www.ams.org/journals/bull/1946-52-04/S0002-9904-1946-08560-2/home.html}.

\bibitem[EMMZ06]{Eckmann2006}
Jean-Pierre Eckmann, Carlos Mej\'{\i}a-Monasterio, and Emmanuel Zabey.
\newblock Memory effects in nonequilibrium transport for deterministic
  {H}amiltonian systems.
\newblock {\em Journal of Statistical Physics}, 123(6):1339--1360, June 2006.
\newblock \url{https://link.springer.com/article/10.1007/s10955-006-9153-4}.

\bibitem[ER96]{Earman1996}
John Earman and Mikl\'os R\'edei.
\newblock Why ergodic theory does not explain the success of equilibrium
  statistical mechanics.
\newblock {\em The British Journal for the Philosophy of Science}, 47(1):63,
  1996.

\bibitem[ET90]{ET90}
L\'aszl\'o Erd{\H o}s and Dao~Quang Tuyen.
\newblock Ergodic properties of the multidimensional {R}ayleigh gas with a
  semipermeable barrier.
\newblock {\em Journal of Statistical Physics}, 59(5):1589--1602, June 1990.
\newblock \url{https://link.springer.com/article/10.1007/BF01334766}.

\bibitem[EY06]{EY06}
Jean-Pierre Eckmann and Lai-Sang Young.
\newblock Nonequilibrium energy profiles for a class of 1-d models.
\newblock {\em Communications in Mathematical Physics}, 262(1):237--267, 2006.
\newblock \url{https://link.springer.com/article/10.1007/s00220-005-1462-y}.

\bibitem[FU14]{FU14}
Krzysztof Fraczek and Corinna Ulcigrai.
\newblock Non-ergodic $\mathbb{Z}$-periodic billiards and infinite translation
  surfaces.
\newblock {\em Inventiones mathematicae}, 197(2):241--298, 2014.
\newblock \url{https://link.springer.com/article/10.1007/s00222-013-0482-z}.

\bibitem[Gal75]{G75}
Giovanni Gallavotti.
\newblock Lectures on the billiard.
\newblock In J{\"u}rgen~K Moser, editor, {\em Dynamical systems, theory and
  applications}, volume~38 of {\em Lecture Notes in Physics}, pages 236--295.
  Springer, Berlin Heidelberg, 1975.
\newblock \url{http://link.springer.com/10.1007/3-540-07171-7_7}.

\bibitem[Gas92]{Gaspard1992}
Pierre Gaspard.
\newblock Diffusion, effusion, and chaotic scattering: An exactly solvable
  liouvillian dynamics.
\newblock {\em Journal of Statistical Physics}, 68(5):673--747, 1992.
\newblock \url{https://link.springer.com/article/10.1007/BF01048873}.

\bibitem[Gas98]{G98}
Pierre Gaspard.
\newblock {\em Chaos, Scattering and Statistical Mechanics}.
\newblock Cambridge Nonlinear Science Series. Cambridge University Press,
  Cambridge, UK, 1998.
\newblock
  \url{https://www.cambridge.org/core/books/chaos-scattering-and-statistical-mechanics/B2F337764BE796C22EE7272DF78DF7F6}.

\bibitem[GG08a]{Gaspard2008}
Pierre Gaspard and Thomas Gilbert.
\newblock Heat conduction and {F}ourier's law by consecutive local mixing and
  thermalization.
\newblock {\em Physical Review Letters}, 101(2):20601, July 2008.
\newblock
  \url{https://journals.aps.org/prl/abstract/10.1103/PhysRevLett.101.020601}.

\bibitem[GG08b]{GG08}
Pierre Gaspard and Thomas Gilbert.
\newblock Heat conduction and {F}ourier's law in a class of many particle
  dispersing billiards.
\newblock {\em New Journal of Physics}, 10:103004, 2008.
\newblock
  \url{https://iopscience.iop.org/article/10.1088/1367-2630/10/10/103004}.

\bibitem[GG12]{Gaspard2012}
Pierre Gaspard and Thomas Gilbert.
\newblock A two-stage approach to relaxation in billiard systems of locally
  confined hard spheres.
\newblock {\em Chaos}, 22:026117, jun 2012.
\newblock \url{https://aip.scitation.org/doi/10.1063/1.3697689}.

\bibitem[GG17]{GG17}
Pierre Gaspard and Thomas Gilbert.
\newblock Dynamical contribution to the heat conductivity in stochastic energy
  exchanges of locally confined gases.
\newblock {\em Journal of Statistical Mechanics: Theory and Experiment}, page
  043210, 2017.
\newblock \url{https://iopscience.iop.org/article/10.1088/1742-5468/aa60ce}.

\bibitem[GGMMS20]{GGM-MSzXX}
Pierre Gaspard, Thomas Gilbert, Carlos Mej\'ia-Monasterio, and Domokos Sz\'asz.
\newblock Billiards on helicoidal surfaces.
\newblock Unpublished, 2020.

\bibitem[GLA75]{GLA75}
Sheldon Goldstein, Joel~L Lebowitz, and Michael Aizenman.
\newblock Ergodic properties of infinite systems.
\newblock In J{\"u}rgen~K Moser, editor, {\em Dynamical systems, theory and
  applications}, volume~38 of {\em Lecture Notes in Physics}, pages 112--143.
  Springer, Berlin Heidelberg, 1975.
\newblock \url{https://link.springer.com/chapter/10.1007/3-540-07171-7_2}.

\bibitem[GLR82]{GLR82}
Sheldon Goldstein, Joel~L Lebowitz, and Krishnamurthi Ravishankar.
\newblock Ergodic properties of a system in contact with a heat bath: A one
  dimensional model.
\newblock {\em Communications in Mathematical Physics}, 85:419, September 1982.
\newblock \url{https://link.springer.com/article/10.1007/BF01208722}.

\bibitem[GO74]{GO74}
Giovanni Gallavotti and Donald~S Ornstein.
\newblock Billiards and {B}ernoulli schemes.
\newblock {\em Communications in Mathematical Physics}, 38(2):83--101, June
  1974.
\newblock \url{https://link.springer.com/article/10.1007/BF01651505}.

\bibitem[{Gra}87]{Gr87}
Walter~T {Grandy Jr}.
\newblock {\em Foundations of statistical mechanics: Equilibrium theory},
  volume~19.
\newblock D. Reidel Publishing Company, Dordrecht, Holland, 1987.
\newblock \url{https://link.springer.com/book/10.1007/978-94-009-3867-0}.

\bibitem[Hed39]{He39}
Gustav~A Hedlund.
\newblock The dynamics of geodesic flows.
\newblock {\em Bulletin of the American Mathematical Society}, 45(4):241--260,
  1939.
\newblock
  \url{https://www.ams.org/journals/bull/1939-45-04/S0002-9904-1939-06945-0/home.html}.

\bibitem[Hof06]{Hoffmann2006}
Dieter Hoffmann.
\newblock Paul drude (1863--1906).
\newblock {\em Annalen der Physik}, 15(7‐8):449--460, 2006.
\newblock \url{https://onlinelibrary.wiley.com/doi/abs/10.1002/andp.200610210}.

\bibitem[Hop37]{Hopf1937}
Eberhard Hopf.
\newblock {\em Ergodentheorie}.
\newblock Springer, Berlin Heidelberg, 1937.
\newblock \url{http://link.springer.com/10.1007/978-3-642-86630-2}.

\bibitem[Hop39]{Ho39}
Eberhard Hopf.
\newblock Statistik der geod{\"a}tischen linien in mannigfaltigkeiten negativer
  kr{\"u}mmung.
\newblock {\em Berichte {\"u}ber die Verhandlungen der S{\"a}chsischen Akademie
  der Wissenschaften zu Leipzig Mathematisch--Physikalische Klasse},
  91:261--304, 1939.

\bibitem[HP19]{A19}
Helge Holden and Ragni Piene, editors.
\newblock {\em The Abel Prize 2013--2017}, Cham, 2019. Norwegian Academy of
  Science and Letters, Springer.
\newblock \url{https://link.springer.com/book/10.1007/978-3-319-99028-6}.

\bibitem[HSV99]{Hirata1999}
Masaki Hirata, Beno{\^\i}t Saussol, and Sandro Vaienti.
\newblock Statistics of return times: A general framework and new applications.
\newblock {\em Communications in Mathematical Physics}, 206(1):33--55, 1999.
\newblock \url{https://link.springer.com/article/10.1007/s002200050697}.

\bibitem[HW80]{Hardy1980}
Jean Hardy and Jaques Weber.
\newblock Diffusion in a periodic wind--tree model.
\newblock {\em Journal of Mathematical Physics}, 21(7):1802--1808, 1980.
\newblock \url{https://aip.scitation.org/doi/10.1063/1.524633}.

\bibitem[HZ00]{H00}
Martin Henk and Chuanming Zong.
\newblock Segments in ball packings.
\newblock {\em Mathematika}, 47(1--2):31--38, 2000.
\newblock
  \url{https://www.cambridge.org/core/journals/mathematika/article/segments-in-ball-packings/43AD62F49B0308E16D18FFD05D3D471B}.

\bibitem[Kac47]{Kac1947}
Mark Kac.
\newblock Random walk and the theory of brownian motion.
\newblock {\em The American Mathematical Monthly}, 54(7):369--391, August 1947.
\newblock
  \url{https://www.tandfonline.com/doi/full/10.1080/00029890.1947.11990189}.

\bibitem[Kar07]{K07}
Mehran Kardar.
\newblock {\em Statistical Physics of Particles}.
\newblock Cambridge University Press, Cambridge, 2007.
\newblock \url{http://ebooks.cambridge.org/ref/id/CBO9780511815898}.

\bibitem[Kel77]{K77}
Gerhard Keller.
\newblock Diplomarbeit.
\newblock Master's thesis, Universit\"at Erlangen, 1977.

\bibitem[KM81]{KM81}
Izumi Kubo and Hiroshi Murata.
\newblock Perturbed billiard systems, {II.} {B}ernoulli properties.
\newblock {\em Nagoya Mathematical Journal}, 81:1–25, 1981.
\newblock
  \url{https://www.cambridge.org/core/product/identifier/S0027763000019127/type/journal_article}.

\bibitem[Kna87]{K87}
Andreas Knauf.
\newblock Ergodic and topological properties of coulombic periodic potentials.
\newblock {\em Communications in Mathematical Physics}, 110:89, March 1987.
\newblock \url{https://link.springer.com/article/10.1007/BF01209018}.

\bibitem[Kry80]{K50}
Nikolai~S Krylov.
\newblock {\em Works on the Foundations of Statistical Physics}.
\newblock Princeton University Press, Princeton, NJ, 1980.
\newblock
  \url{https://press.princeton.edu/books/hardcover/9780691643748/works-on-the-foundations-of-statistical-physics},
  original: Publishing House of Soviet Acad Sci, Moscow, 1950.

\bibitem[KSS91]{KSSz91}
Andr{\'a}s Kr{\'a}mli, N{\'a}ndor Sim\'anyi, and Domokos Sz{\'a}sz.
\newblock The k--property of three billiard balls.
\newblock {\em The Annals of Mathematics}, 133(1):37--72, 1991.
\newblock \url{https://annals.math.princeton.edu/articles/14549}.

\bibitem[KSS92]{KSSz92}
Andr{\'a}s Kr{\'a}mli, N{\'a}ndor Sim{\'a}nyi, and Domokos Sz{\'a}sz.
\newblock The k--property of four billiard balls.
\newblock {\em Communications in Mathematical Physics}, 144(1):107--148,
  February 1992.
\newblock \url{https://link.springer.com/article/10.1007/BF02099193}.

\bibitem[Kub76]{K76}
Izumi Kubo.
\newblock Perturbed billiard systems, {I.} {T}he ergodicity of the motion of a
  particle in a compound central field.
\newblock {\em Nagoya Mathematical Journal}, 61:1–57, 1976.
\newblock
  \url{https://www.cambridge.org/core/product/identifier/S0027763000017281/type/journal_article}.

\bibitem[Lep16]{Lepri2016}
Stefano Lepri, editor.
\newblock {\em Thermal Transport in Low Dimensions: From Statistical Physics to
  Nanoscale Heat Transfer}, Cham, 2016. Springer International Publishing.
\newblock \url{https://link.springer.com/book/10.1007/978-3-319-29261-8}.

\bibitem[Liv19]{L18}
Carlangelo Liverani.
\newblock Transport in partially hyperbolic fast--slow systems.
\newblock In Boyan Sirakov, {de Souza}~Paulo Ney, and Viana Marcelo, editors,
  {\em Proceedings of the International Congress of Mathematicians 2018}, pages
  2643--2667, Singapore, June 2019. World Scientific.
\newblock
  \url{https://www.worldscientific.com/doi/abs/10.1142/9789813272880_0154}.

\bibitem[LL80]{LL}
Lev~D Landau and Evgeny~M Lifshitz.
\newblock {\em Statistical Physics}, volume~5 of {\em Part I}.
\newblock Pergamon Press, 3rd edition, 1980.
\newblock
  \url{https://www.sciencedirect.com/book/9780080570464/statistical-physics}.

\bibitem[LLMM03]{LLM-M03}
Hern{\'a}n Larralde, Fran{\c c}ois Leyvraz, and Carlos Mej{\'i}a~Monasterio.
\newblock Transport properties of a modified {L}orentz gas.
\newblock {\em Journal of Statistical Physics}, 113(1):197--231, 2003.
\newblock \url{https://link.springer.com/article/10.1023/A:1025726905782}.

\bibitem[LLP03]{LLP03}
Stefano Lepri, Roberto Livi, and Antonio Politi.
\newblock Thermal conduction in classical low--dimensional lattices.
\newblock {\em Physics Reports}, 377:1, April 2003.
\newblock
  \url{https://www.sciencedirect.com/science/article/abs/pii/S0370157302005586}.

\bibitem[Lor05]{L05}
Hendrik~A Lorentz.
\newblock The motion of electrons in metallic bodies i.
\newblock {\em Koninklijke Nederlandse Akademie van Wetenschappen Proceedings},
  7:438--453, 1905.
\newblock \url{https://www.dwc.knaw.nl/DL/publications/PU00013989.pdf}.

\bibitem[LP73]{Lebowitz1973}
Joel~L Lebowitz and Oliver Penrose.
\newblock Modern ergodic theory.
\newblock {\em Physics Today}, 26(2):23--29, 1973.
\newblock \url{http://physicstoday.scitation.org/doi/10.1063/1.3127948}.

\bibitem[LT19]{LT19}
Christopher Lutsko and B{\'a}lint T{\'o}th.
\newblock Invariance principle for the random wind--tree process.
\newblock ar{X}iv:1912.02492, dec 2019.
\newblock \url{https://arxiv.org/abs/1912.02492}.

\bibitem[LT20]{LT18}
Christopher Lutsko and B{\'a}lint T{\'o}th.
\newblock Invariance principle for the random lorentz gas—{Beyond} the
  {Boltzmann}-{Grad} limit.
\newblock {\em Communications in Mathematical Physics}, 2020.
\newblock \url{https://doi.org/10.1007/s00220-020-03852-8}.

\bibitem[Mar14]{Marklof2014}
Jens Marklof.
\newblock The low-density limit of the {L}orentz gas: periodic, aperiodic and
  random.
\newblock In {\em Proceedings of the International Congress of Mathematicians},
  volume {III}, pages 623--646. International Mathematical Union, Seoul, 2014.
\newblock
  \url{https://www.mathunion.org/fileadmin/ICM/Proceedings/ICM2014.3/ICM2014.3.pdf}.

\bibitem[Men19]{Mendl2019}
Christian~B. Mendl.
\newblock Fourier's law and many-body quantum systems.
\newblock {\em Comptes Rendus Physique}, 20(5):442--448, 2019.
\newblock
  \url{http://www.sciencedirect.com/science/article/pii/S1631070519301124}.

\bibitem[MM74]{MM74}
Lawrence Markus and Kenneth~R Meyer.
\newblock {\em Generic {H}amiltonian dynamical systems are neither integrable
  nor ergodic}.
\newblock 144. Memoirs of the American Mathematical Society, Providence, RI,
  1974.
\newblock \url{https://bookstore.ams.org/memo-1-144/}.

\bibitem[MMLL01]{M-MCL01}
Carlos Mej\'{\i}a-Monasterio, Hern{\'a}n Larralde, and Fran{\c{c}}ois Leyvraz.
\newblock Coupled normal heat and matter transport in a simple model system.
\newblock {\em Physical Review Letters}, 86:5417--5420, 2001.
\newblock \url{https://link.aps.org/doi/10.1103/PhysRevLett.86.5417}.

\bibitem[Moo15]{Moore2015}
Calvin~C Moore.
\newblock Ergodic theorem, ergodic theory, and statistical mechanics.
\newblock {\em Proceedings of the National Academy of Sciences},
  112(7):1907--1911, 2015.
\newblock \url{https://www.pnas.org/content/112/7/1907}.

\bibitem[MS19]{MS19}
Jens Marklof and Andreas Str{\"o}mbergsson.
\newblock Kinetic theory for the low--density {L}orentz gas.
\newblock ar{X}iv:1910.04982, October 2019.
\newblock \url{https://arxiv.org/abs/1910.04982}.

\bibitem[MT16]{M-ST16}
Alba {M\'alaga Sabogal} and Serge Troubetzkoy.
\newblock Ergodicity of the ehrenfest wind--tree model.
\newblock {\em Comptes Rendus Mathematique}, 354(10):1032--1036, 2016.
\newblock
  \url{http://www.sciencedirect.com/science/article/pii/S1631073X16301595}.

\bibitem[MZ83]{MZ1983}
Jonathan Machta and Robert Zwanzig.
\newblock Diffusion in a periodic {L}orentz gas.
\newblock {\em Physical Review Letters}, 50(25):1959--1962, June 1983.
\newblock
  \url{https://journals.aps.org/prl/abstract/10.1103/PhysRevLett.50.1959}.

\bibitem[Nar99]{N99}
Thiruppudaimarudhur~N Narasimhan.
\newblock Fourier's heat conduction equation: History, influence, and
  connections.
\newblock {\em Reviews of Geophysics}, 37(1):151--172, 1999.
\newblock
  \url{https://agupubs.onlinelibrary.wiley.com/doi/abs/10.1029/1998RG900006}.

\bibitem[NSV14]{N14}
P{\'e}ter N{\'a}ndori, Domokos Sz{\'a}sz, and Tam{\'a}s Varj{\'u}.
\newblock Tail asymptotics of free path lengths for the periodic lorentz
  process. on dettmann's geometric conjectures.
\newblock {\em Communications in Mathematical Physics}, 331:111–137, 2014.
\newblock \url{https://link.springer.com/article/10.1007/s00220-014-2086-x}.

\bibitem[P{\`e}n09]{Pene2009}
Françoise P{\`e}ne.
\newblock Asymptotic of the number of obstacles visited by the planar lorentz
  process.
\newblock {\em Discrete \& Continuous Dynamical Systems - A}, 24:567--587,
  2009.
\newblock
  \url{http://aimsciences.org//article/id/2764fe79-6383-459c-9817-430529287fab}.

\bibitem[P{\`e}n19]{Pene2019}
Françoise P{\`e}ne.
\newblock Mixing and decorrelation in infinite measure: The case of the
  periodic sinai billiard.
\newblock {\em Annales de l'Institut Henri Poincaré, Probabilités et
  Statistiques}, 55(1):378--411, 02 2019.
\newblock \url{https://projecteuclid.org/euclid.aihp/1547802404}.

\bibitem[Per09]{Perrin1909}
Jean~Baptiste Perrin.
\newblock Mouvement brownien et r{\'e}alit{\'e} mol{\'e}culaire.
\newblock {\em Annales de chimie et de physique}, 18(8):5--114, 1909.
\newblock \url{https://gallica.bnf.fr/ark:/12148/bpt6k349481?rk=64378}.

\bibitem[PGS12]{PajorGyulai2012}
Zsolt Pajor-Gyulai and Domokos Sz{\'a}sz.
\newblock Energy transfer and joint diffusion.
\newblock {\em Journal of Statistical Physics}, 146(5):1001--1025, January
  2012.
\newblock \url{https://link.springer.com/article/10.1007/s10955-012-0426-9}.

\bibitem[PGST10]{PajorGyulai2010}
Zsolt Pajor-Gyulai, Domokos Sz{\'a}sz, and Imre~P{\'e}ter T{\'o}th.
\newblock Billiard models and energy transfer.
\newblock In Pavel Exner, editor, {\em XVIth International Congress on
  Mathematical Physics. Prague. Held 3-8 August 2009}, pages 328--332. World
  Scientific, 2010.
\newblock \url{http://www.worldscientific.com/worldscibooks/10.1142/7727}.

\bibitem[Pro56]{P56}
Yuri~V Prokhorov.
\newblock Convergence of random processes and limit theorems in probability
  theory.
\newblock {\em Theory of Probability \& Its Applications}, 1(2):157--214, 1956.
\newblock \url{https://epubs.siam.org/doi/10.1137/1101016}.

\bibitem[PS10]{Pene2010}
Fran{\c c}oise P{\`e}ne and Beno{\^\i}t Saussol.
\newblock Back to balls in billiards.
\newblock {\em Communications in Mathematical Physics}, 293:837--866, 2010.
\newblock \url{https://link.springer.com/article/10.1007/s00220-009-0911-4}.

\bibitem[RKN00]{Rateitschak2000}
Katja Rateitschak, Rainer Klages, and Gregoire Nicolis.
\newblock Thermostating by deterministic scattering: the periodic lorentz gas.
\newblock {\em Journal of Statistical Physics}, 99(5--6):1339--1364, 2000.
\newblock \url{https://link.springer.com/article/10.1023/A:1018645007533}.

\bibitem[San08]{S08}
David~P Sanders.
\newblock Normal diffusion in crystal structures and higher-dimensional
  billiard models with gaps.
\newblock {\em Physical Review E}, 78:060101, 2008.
\newblock
  \url{https://journals.aps.org/pre/abstract/10.1103/PhysRevE.78.060101}.

\bibitem[Sas15]{Sasada2015}
Makiko Sasada.
\newblock Spectral gap for stochastic energy exchange model with nonuniformly
  positive rate function.
\newblock {\em Annals of Probability}, 43(4):1663--1711, 07 2015.
\newblock \url{https://projecteuclid.org/euclid.aop/1433341317}.

\bibitem[SC87]{SCh87}
Yakov~G Sinai and Nikolai Chernov.
\newblock Ergodic properties of certain systems of two--dimensional discs and
  three-dimensional balls.
\newblock {\em Russian Mathematical Surveys}, 42:181--207, June 1987.
\newblock
  \url{https://iopscience.iop.org/article/10.1070/RM1987v042n03ABEH001421}.

\bibitem[Sim13]{S13}
N{\'a}ndor Sim{\'a}nyi.
\newblock Singularities and non--hyperbolic manifolds do not coincide.
\newblock {\em Nonlinearity}, 26(6):1703--1717, June 2013.
\newblock \url{http://iopscience.iop.org/article/10.1088/0951-7715/26/6/1703}.

\bibitem[Sim19]{S19}
N{\'a}ndor Sim{\'a}nyi.
\newblock Further developments of sinai’s ideas: The boltzmann–sinai
  hypothesis.
\newblock In Helge Holden and Ragni Piene, editors, {\em The Abel Prize
  2013--2017}, pages 287--298. Springer, Cham, 2019.
\newblock \url{https://link.springer.com/chapter/10.1007/978-3-319-99028-6_12}.

\bibitem[Sin60]{S60}
Yakov~G Sinai.
\newblock Geodesic flows on manifolds of negative constant curvature.
\newblock {\em Doklady Akademii Nauk SSSR}, 131(4):752--755, 1960.
\newblock \url{http://mi.mathnet.ru/eng/dan39938}.

\bibitem[Sin63]{S63}
Yakov~G Sinai.
\newblock On the foundations of the ergodic hypothesis for a dynamical system
  of statistical mechanics.
\newblock {\em Doklady Akademii Nauk SSSR}, 153(6):1261--1264, 1963.
\newblock \url{http://mi.mathnet.ru/eng/dan28929}.

\bibitem[Sin70]{S70}
Yakov~G Sinai.
\newblock Dynamical systems with elastic reflections: Ergodic properties of
  dispersing billiards.
\newblock {\em Russian Mathematical Surveys}, 25:137--189, 1970.
\newblock
  \url{https://iopscience.iop.org/article/10.1070/RM1970v025n02ABEH003794/meta}.

\bibitem[Sin79]{S79}
Yakov~G Sinai.
\newblock Ergodic properties of the {L}orentz gas.
\newblock {\em Functional Analysis and Its Applications}, 13(3):192--202, July
  1979.
\newblock \url{https://link.springer.com/article/10.1007/BF01077487}.

\bibitem[Sma80]{Smale1980}
Steve Smale.
\newblock {\em The Mathematics of Time: Essays on Dynamical Systems, Economic
  Processes, and Related Topics}.
\newblock Springer, New York, NY, 1980.
\newblock \url{https://link.springer.com/book/10.1007/978-1-4613-8101-3}.

\bibitem[Spo91]{S91}
Herbert Spohn.
\newblock {\em Large Scale Dynamics of Interacting Particles}.
\newblock Springer, Berlin Heidelberg, 1991.
\newblock \url{http://link.springer.com/book/10.1007/978-3-642-84371-6}.

\bibitem[SS86]{SS86}
Yakov~G Sinai and M.~R. Soloveichik.
\newblock One-dimensional classical massive particle in the ideal gas.
\newblock {\em Communications in Mathematical Physics}, 104:423--443, 1986.
\newblock \url{https://link.springer.com/article/10.1007/BF01210949}.

\bibitem[ST86]{SzT86}
Domokos Sz\'asz and B\'alint T\'oth.
\newblock Bounds for the limiting variance of the ``heavy particle'' in
  $\mathbb{R}^{1}$.
\newblock {\em Communications in Mathematical Physics}, 104:445--455, 1986.
\newblock \url{https://link.springer.com/article/10.1007/BF01210950}.

\bibitem[ST87]{SzT87}
Domokos Sz{\'a}sz and Balint T{\'o}th.
\newblock Towards a unified dynamical theory of the brownian particle in an
  ideal gas.
\newblock {\em Communications in Mathematical Physics}, 111:41--62, 1987.
\newblock \url{https://link.springer.com/article/10.1007/BF01239014}.

\bibitem[Str91]{Rayleigh1891}
{Lord Rayleigh} (J~W Strutt).
\newblock Dynamical problems in illustration of the theory of gases.
\newblock {\em The London, Edinburgh, and Dublin Philosophical Magazine and
  Journal of Science}, 32(198):424--445, 1891.
\newblock \url{https://www.tandfonline.com/doi/abs/10.1080/14786449108620207}.

\bibitem[Sz{\'a}00]{Sz96}
Domokos Sz{\'a}sz.
\newblock Boltzmann's ergodic hypothesis, a conjecture for centuries.
\newblock In Domokos Sz{\'a}sz, editor, {\em Hard Ball Systems and the
  {L}orentz Gas}, volume 101 of {\em Encyclopaedia of Mathematical Sciences},
  pages 421--446. Springer, Berlin, Heidelberg, 2000.
\newblock \url{https://link.springer.com/chapter/10.1007/978-3-662-04062-1_14},
  Reprinted from Studia Scientiarum Hungarica \textbf{31} 266--322 (1996).

\bibitem[Sz{\'a}14]{Sz14}
Domokos Sz{\'a}sz.
\newblock Mathematical billiards and chaos.
\newblock {\em Newsletter of the European Matheamtical Society}, 93:22--29,
  2014.
\newblock \url{https://www.ems-ph.org/journals/newsletter/pdf/2014-09-93.pdf},
  Lecture recording \url{https://www.abelprize.no/artikkel/vis.html?tid=61307}.

\bibitem[Sz{\'a}19]{Sz19}
Domokos Sz{\'a}sz.
\newblock Markov approximations and statistical properties of billiards.
\newblock In Helge Holden and Ragni Piene, editors, {\em The Abel Prize
  2013--2017}, pages 299--319. Springer, Cham, 2019.
\newblock \url{https://link.springer.com/chapter/10.1007/978-3-319-99028-6_13}.

\bibitem[TKS92]{Toda1992}
Morikazu Toda, Ryogo Kubo, and Nobuhiko Sait{\^o}.
\newblock {\em Statistical Physics I: Equilibrium Statistical Mechanics}.
\newblock Springer, Berlin, Heidelberg, 1992.
\newblock \url{https://link.springer.com/book/10.1007/978-3-642-58134-2}.

\bibitem[Var93]{Varadhan1993}
S~R~Srinivasa Varadhan.
\newblock Nonlinear diffusion limit for a system with nearest neighbor
  interactions. {II}.
\newblock In K~David Elworthy and Nobuyuki Ikeda, editors, {\em Asymptotic
  problems in probability theory: stochastic models and diffusions on
  fractals}, number 283 in Pitman Research Notes in Mathematics, pages 75--128.
  Longman Scientific \& Technical, Essex, England, 1993.
\newblock "Proceedings of the Taniguchi international symposium, Sanda and
  Kyoto, 1990.".

\bibitem[Vet84]{V84}
Andr\'as Vetier.
\newblock Sinai--billiard in potential field (absolute continuity).
\newblock In J~Mogyoródi, I~Vincze, and W~Wertz, editors, {\em Statistics and
  Probability (Proceedings of the 3rd Pannonian Symposium on Mathematical
  Statistics, Visegr\'ad, Hungary, 13--18 September 1982)}, pages 341--351. D
  Reidel Publishing Company, Dordrecht, 1984.

\bibitem[Vet89]{V89}
Andr{\'a}s Vetier.
\newblock Sinai--billiard in potential field (ergodic components).
\newblock {\em Banach Center Publications}, 23(1):313--326, 1989.
\newblock \url{http://www.impan.pl/get/doi/10.4064/-23-1-313-326}.

\bibitem[VG03]{V03}
S\'ebastien Viscardy and Pierre Gaspard.
\newblock Viscosity in molecular dynamics with periodic boundary conditions.
\newblock {\em Physical Review E}, 68(4):041204, October 2003.
\newblock
  \url{https://journals.aps.org/pre/abstract/10.1103/PhysRevE.68.041204}.

\bibitem[{von}32]{N32}
John {von Neumann}.
\newblock Proof of the quasi--ergodic hypothesis.
\newblock {\em Proceedings of the National Academy of Sciences}, 18(1):70--82,
  1932.
\newblock \url{https://www.pnas.org/content/18/1/70}.

\bibitem[VS71]{VS71}
Kerim~L Volkovyskii and Yakov~G Sinai.
\newblock Ergodic properties of an ideal gas with an infinite number of degrees
  of freedom.
\newblock {\em Functional Analysis and Its Applications}, 5(3):185--187, July
  1971.
\newblock \url{https://link.springer.com/article/10.1007/BF01078123}.

\bibitem[VW85]{VW85}
Giorgio Velo and Arthur~S Wightman, editors.
\newblock {\em Regular and Chaotic Motions in Dynamic Systems}, volume 118 of
  {\em NATO ASI Series}, New York, NY, 1985. Plenum Press.
\newblock \url{http://link.springer.com/10.1007/978-1-4684-1221-5}.

\bibitem[Wie23]{Wiener1923}
Norbert Wiener.
\newblock Differential--space.
\newblock {\em Journal of Mathematics and Physics}, 2(1--4):131--174, 1923.
\newblock \url{https://onlinelibrary.wiley.com/doi/abs/10.1002/sapm192321131}.

\bibitem[Wig60]{W60}
Eugene~P Wigner.
\newblock The unreasonable effectiveness of mathematics in the natural
  sciences.
\newblock {\em Communications on Pure and Applied Mathematics}, 13(1):1--14,
  1960.
\newblock \url{https://onlinelibrary.wiley.com/doi/abs/10.1002/cpa.3160130102}.

\bibitem[You98]{Y98}
Lai-Sang Young.
\newblock Statistical properties of dynamical systems with some hyperbolicity.
\newblock {\em Annals of Mathematics}, 147:585--650, 1998.
\newblock \url{http://annals.math.princeton.edu/articles/12940}.

\bibitem[Zor06]{Zorich2006}
Anton Zorich.
\newblock Flat surfaces.
\newblock In Pierre Cartier, Bernard Julia, Pierre Moussa, and Pierre Vanhove,
  editors, {\em Frontiers in Number Theory, Physics, and Geometry I},
  chapter~13, pages 439--585. Springer, Berlin--Heidelberg, 2006.
\newblock \url{https://www.springer.com/gp/book/9783540231899}.

\end{thebibliography}
\newcommand{\etalchar}[1]{$^{#1}$}

\end{document}